\documentclass[11pt, twoside]{article}
\pdfoutput=1
\usepackage{a4wide,amssymb,cite,graphicx}
\usepackage{epsfig}
\usepackage[usenames,dvipsnames]{color}
\usepackage{slashed}

\newcommand{\be}{\begin{equation}}
\newcommand{\ee}{\end{equation}}
\newcommand{\bea}{\begin{eqnarray}}
\newcommand{\eea}{\end{eqnarray}}

\begin{document}

\color{black}

\begin{flushright}
CERN-PH-TH/2013-175
\end{flushright}

\vspace{0.5cm}

\begin{center}
{\Large\bf\color{black} Gravity-mediated (or Composite) Dark Matter \\[3mm]  Confronts Astrophysical Data}\\
\bigskip\color{black}\vspace{1.0cm}{
{\bf Hyun Min Lee$^{1,*}$, Myeonghun Park$^{2,3,\dagger}$ and Ver\'onica Sanz$^{4\ddagger}$ }
\vspace{0.5cm}
} \\[8mm]

{\it $^1$Department of Physics, Chung-Ang University, Seoul 156-756, Korea.} \\
{\it $^2$ Kavli Institute for the Physics and Mathematics of the Universe (WPI), Todai Institutes for Advanced Study, The University of Tokyo, Japan.}\\
{\it $^3$ Theory Division, Physics Department, CERN,  CH--1211 Geneva 23,  Switzerland.}\\
{\it $^4$Department of Physics and Astronomy, University of Sussex, Brighton BN1 9QH, UK.  } \\
\end{center}
\bigskip
\centerline{\bf Abstract}
\begin{quote}
We consider the astrophysical bounds on a new form of dark matter, the so called Gravity-mediated Dark Matter. In this scenario, dark matter communicates with us through a mediator sector composed of gravitational resonances, namely a new scalar (radion) and a massive spin-two resonance (massive graviton). We consider specific models motivated by natural electroweak symmetry breaking or weak-scale dark matter in the context of models in warped extra-dimensions and their composite duals. The main Dark Matter annihilation mechanism is due to the interactions of KK gravitons to gauge bosons that propagate in bulk. 
We impose the bounds on monochromatic or continuum photons from Fermi-LAT and HESS.
We also explore scenarios in which the Fermi gamma-ray line could be a manifestation of Gravity-mediated Dark Matter. 
\end{quote}

\vspace{2cm}

\begin{flushleft}
$^*$Email: hminlee@cau.ac.kr \\
$^\dagger$Email: myeonghun.park@cern.ch  \\
$^\ddagger$Email: v.sanz@sussex.ac.uk
\end{flushleft}

\thispagestyle{empty}

\normalsize

\newpage

\setcounter{page}{1}

\section{Introduction}

Indirect detection experiments with cosmic rays are a testing ground for the signatures of dark matter(DM) annihilation or decay from the gamma-ray and anti-particle production. In particular, in the case of dark matter annihilation, Weakly Interacting Massive Particles(WIMPs) can provide easily the necessary thermal cross section for the relic density of about $\langle \sigma v\rangle\simeq 3\times 10^{-26}{\rm cm}^3/{\rm s}$ from the thermal freezeout and can be tested by complementarity between direct and indirect detection experiments.
In order for indirect detection to be relevant, dark matter annihilation should not depend on the temperature much, namely, showing the s-wave behavior. 

Recently there has been an interesting indication for dark matter from the gamma-ray line at about $130\,{\rm GeV}$ coming from the galactic center in Fermi-LAT data \cite{weniger,fermi-line,fermi-lat}. The significance of the gamma-ray line has become smaller with reprocessed data than the previous result to $3.3\sigma$, being reduced to $1.6$ at the global level \cite{fermi-lat}.  Nonetheless, it remains to be seen, how the gamma-ray line signal evolves in a near future.
The observed Fermi gamma-ray line implies that the annihilation cross section of dark matter into monochromatic photons is $\langle \sigma v\rangle=(1.27-2.27)\times 10^{-27}{\rm cm}^3/{\rm s}$ \cite{weniger}, depending on whether dark matter profile is given by Einasto or NFW.  But, the corresponding process is  loop-suppressed as compared to the tree-level processes, so  it is challenging to build a microscopic model for dark matter interactions \cite{models,lpp,lps}. Furthermore, the tree-level DM annihilations have been strongly constrained by the gamma-ray searches \cite{fermilat2,hess2013,dwarfgalaxy,continuum,gc} and anti-proton bounds\cite{antiproton,hambye}. Independent of whether the Fermi gamma-ray line survives more data, the gamma-ray constraints are getting more important for dark matter model building in general. 

In this paper, we consider a gravity-mediated dark matter, that has been proposed by the authors \cite{GMDM} to relate dark matter mass to the geometric solution of the hierarchy problem in the 5D warped extra dimension with UV and IR branes \cite{RSmodel}.  We generalize the previous setup to Model A and B, depending on whether the Higgs doublet is localized on IR or UV branes.
In both models, the SM particles propagate in bulk while dark matter of arbitrary spin ($s=0,1/2\,\, {\rm and}\,\, 1$) is localized on the IR brane. When the SM fermions are localized toward the UV brane as is the case for fermion mass hierarchy and natural flavor conservation, dark matter annihilates mainly into gauge bosons living in the bulk.  In Model B,  we show that the Fermi gamma-ray line at about $130\,{\rm GeV}$ can be obtained from the annihilation of a pair of vector dark matter into a photon pair, which is mediated by the KK graviton without a need of large couplings or resonance.
We present the astrophysical bounds from Fermi-LAT and HESS on the models and relate them to the direct detection of dark matter and the discovery potential of the KK graviton at the LHC.

The paper is organized as follows. We begin with the setups for dark matter in the 5D warped spacetime and compute the annihilation cross sections of scalar, fermion and vector dark matter in the models. Then, we search the parameter space for KK graviton coupling and mass, being consistent with the relic density and impose the astrophysical constraints on the model. 
In next section, the direct detection and the collider signatures will be discussed. 
Finally, conclusions are drawn.

\section{Setup}

In this section, we present the general couplings of the KK graviton and the radion to the SM particles, that are determined by the locations of the SM particles in a warped gravitational background with two branes.

\begin{figure}
\begin{minipage}{8in}
\hspace*{-0.7in}
\centerline{\includegraphics[height=6cm]{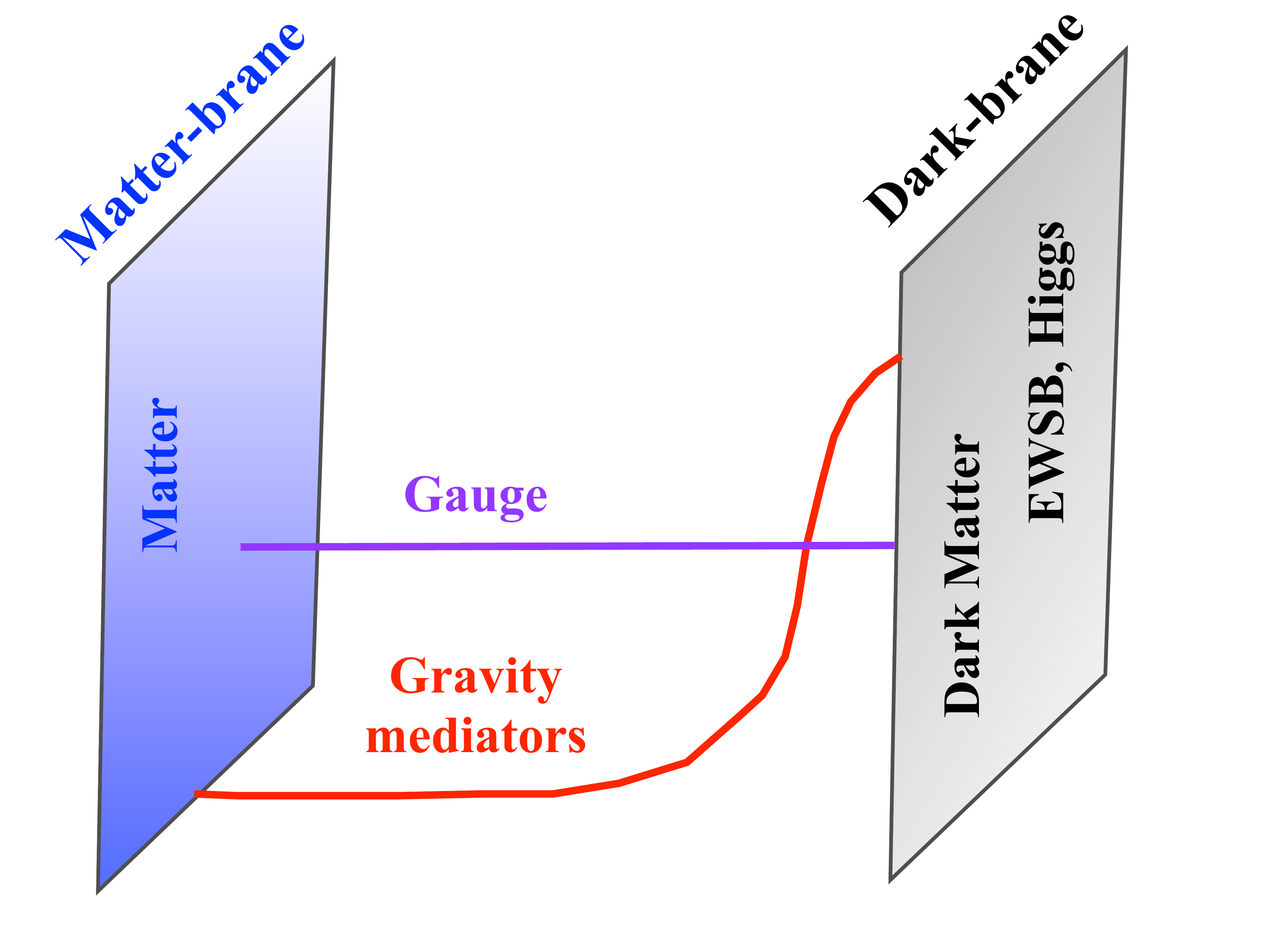} \includegraphics[height=6cm]{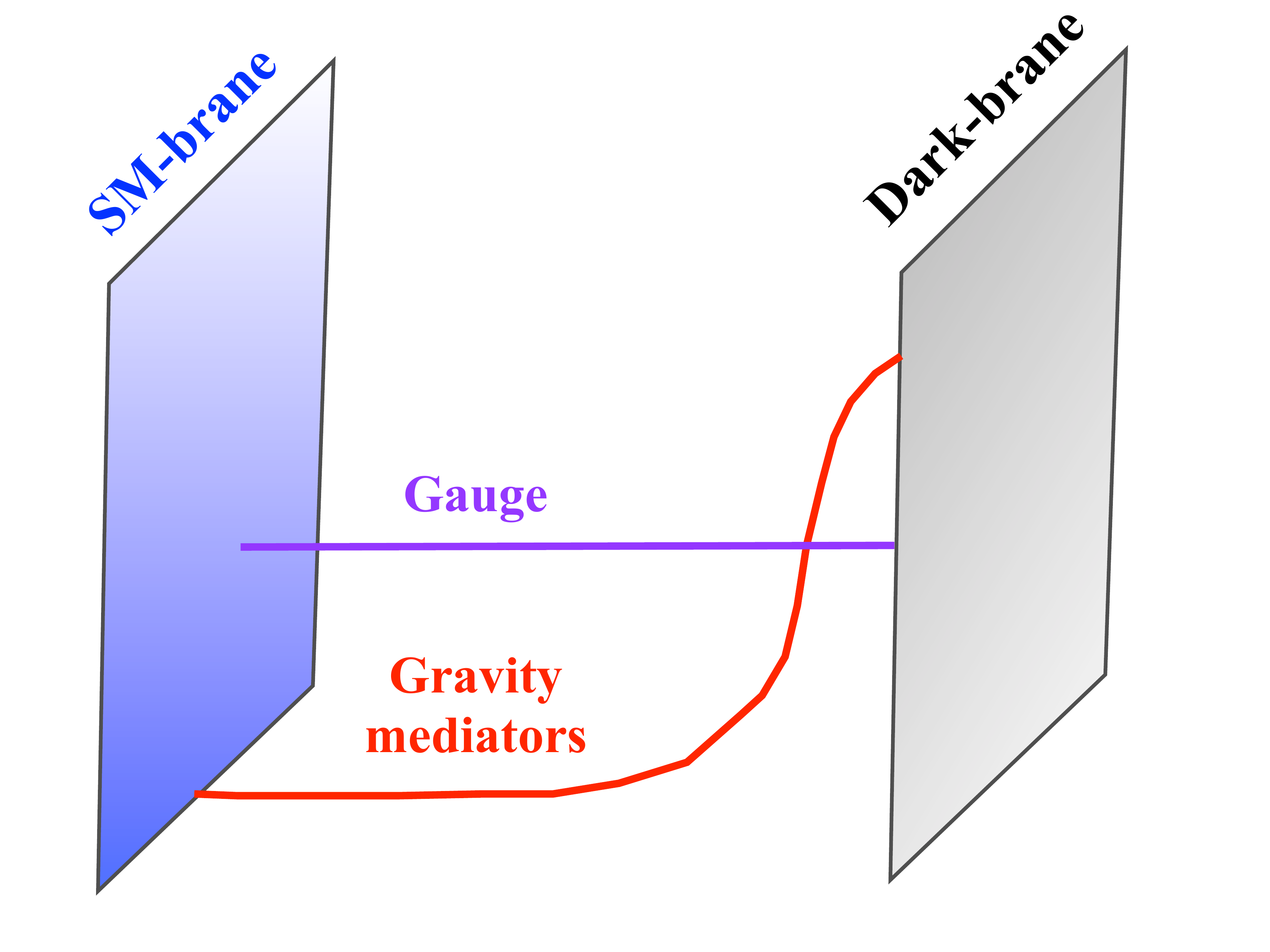}} \label{setupfig}
\hfill
\end{minipage}
\caption{
{\it
Two setups for dark matter in extra-dimensions. Left(Model A): Matter-brane and Dark-brane are at the opposite sides of the extra dimension while Higgs fields are on the Dark-brane. Right(Model B): The Dark-brane and Standard Model brane are at the opposite sides of the extra-dimension. Gauge fields live in the bulk in both cases and Dark Matter communicates to the SM via the gravity mediators.
} 
}
\label{setup}
\end{figure}

There are two setups that are distinguished by the location of the Higgs fields depicted in Fig.~\ref{setup} :

\begin{enumerate}
  \item {\bf Model A - Hierarchy problem: Higgs fields and dark matter are localized on the IR brane while the SM matter is localized on the UV brane.}
  \item  {\bf Model B - WIMP dark matter: dark matter is localized on the IR brane while the SM matter and Higgs fields are localized on the UV brane. }
  \end{enumerate}

In both models, gauge fields are assumed to propagate in bulk. Dark matter has a strong coupling to the KK graviton and the radion, that are localized on the same brane.   
In Model A, dark matter can annihilate dominantly into Higgs degrees of freedom. 
When top quark is localized towards the IR brane, dark matter can annihilate into a top quark pair with sizable branching fraction, if kinematically allowed.
On the other hand, in Model B, the annihilation of dark matter into a pair of the SM matter or Higgs fields is suppressed so dark matter can annihilate dominantly into a pair of SM gauge bosons, leading to a large branching fraction of dark matter annihilation into a photon pair.

In the RS background \cite{RSmodel},  the graviton and the radion are described by the tensor and scalar fluctuations of
the warped metric,
\bea
d s^2= w(z)^2 \left( e^{-2 r} (\eta_{\mu \nu} + G_{\mu\nu} )- (1+ 2 r)^2 d z^2 \right) \ . \label{metric2}
\eea
where $w(z)=1/(k|z|)$ is the warp factor and $G_{\mu\nu}$ and $r$ are 5D fields propagating in the extra dimension.  The fifth dimension is compactified to an interval $z\in[z_0,z_1]$, and four-dimensional branes with nonzero tensions are located at the ends of the extra dimension.

   The present model in extra-dimensions can be interpreted in terms of a strongly-coupled model in four-dimensions. Some details of this duality can be found in Ref.~\cite{GMDM}. 
  
  In a nutshell, the strongly coupled 4D physics evolution from high to low energies is encoded in the values of the parameters when measured at a position $z$ in the extra-dimension. The brane at $z_0$ represents UV boundary conditions to this evolution, and the brane at $z_1$ corresponds to the IR boundary conditions.  Propagation from the Matter towards the Dark brane is equivalent to integrate out degrees of freedom. At a position $z_*$ the {\it local} cutoff is related to the 4D UV cutoff as~\cite{lisa-matt} $\Lambda (z_*) = \omega(z_*) \Lambda_{UV}$. The IR brane encodes information on the physics leading to confinement, and can be used to engineer the spontaneous breaking of a 4D global symmetry due to the strong sector. 
  
 Kaluza-Klein states are the dual of bound states due to confinement physics. Localization near the UV (IR) brane means a small (large) degree of compositeness of the state. De-localized (flat) gauge fields in the extra-dimension represent global symmetries of the composite sector, weakly gauged by the UV dynamics~\cite{csaba-fat}. These flat fields are a mixture of composite and elementary fields, in analogy with the $\rho-\gamma$ mixing in QCD~\cite{ami-qcd,alex-qcd,QCD-hol}.
 
 The presence of gravity mediators is a manifestation of a conformal symmetry of the composite sector, which is spontaneously broken by the strong physics. The radion is dual to the goldstone boson from dilatation symmetry in 4D~\cite{real-radion,Witek-Jiji}, the dilaton.  The dual interpretation of the massive graviton is not so clear. We interpret the massive KK graviton as a manifestation of a CFT diffeomorphism invariance, broken spontaneously by the Dark Brane, but a more rigorous investigation should be done to understand the dual role of the massive graviton. See Ref.~\cite{GMDM} for more details.

\subsection{KK graviton mediator}

We introduce the interactions of the SM particles and dark matter to the KK graviton $G_{\mu\nu}$ as
\bea
{\cal L}_{\rm KK}
&=&-\frac{1}{\Lambda}G^{\mu\nu}\bigg[T^{\rm DM}_{\mu\nu} +c^G_\psi\bigg(\frac{i}{4}{\bar\psi}(\gamma_\mu D_\nu+\gamma_\nu D_\mu)\psi-\frac{i}{4}(D_\mu{\bar\psi}\gamma_\nu+D_\nu{\bar\psi}\gamma_\nu)\psi  \nonumber \\
&&\quad -g_{\mu\nu} ({\bar\psi}\gamma^\mu D_\mu\psi-m_\psi {\bar\psi}\psi)+\frac{i}{2}g_{\mu\nu}\partial^\rho({\bar\psi}\gamma_\rho\psi ) \bigg) \nonumber \\
&&\quad+c^G_V\left(\frac{1}{4}g_{\mu\nu} F^{\lambda\rho}F_{\lambda\rho}-F_{\mu\lambda}F^\lambda\,_\nu\right) \nonumber \\
&&\quad+c^G_H\left(-g_{\mu\nu}D^\rho H^\dagger D_\rho H+g_{\mu\nu}V(H)+D_\mu H^\dagger D_\nu H+D_\nu H^\dagger D_\mu H\right)\bigg]
\eea
with the energy-momentum tensor for dark matter (DM) given by~\footnote{Note that at the level of interactions of the spin-two particle with two SM particles, the structure of the coupling is identical whether it is a massive KK-graviton or a resonance from a new 4D strongly coupled sector. This has been shown in Ref.~\cite{dual-vero}, and it is a consequence of Lorentz, gauge and CP invariance.}
\bea
T^{\rm (Vector~DM)}_{\mu\nu}&=&
 \frac{1}{4}g_{\mu\nu} X^{\lambda\rho}X_{\lambda\rho}-X_{\mu\lambda}X^\lambda\,_\nu+m^2_X\Big(X_\mu X_\nu-\frac{1}{2}g_{\mu\nu} X^\lambda  X_\lambda\Big)  , \nonumber\\
T^{\rm (Fermion~DM)}_{\mu\nu}&=& \frac{i}{4}{\bar\chi}(\gamma_\mu\partial_\nu+\gamma_\nu\partial_\mu)\chi-\frac{i}{4} (\partial_\mu{\bar\chi}\gamma_\nu+\partial_\nu{\bar\chi}\gamma_\nu)\chi-g_{\mu\nu}(i {\bar\chi}\gamma^\mu\partial_\mu\chi- m_\chi {\bar\chi}\chi) 
\nonumber \\
&&+\frac{i}{2}g_{\mu\nu}\partial^\rho({\bar\chi}\gamma_\rho\chi),  \nonumber \\
T^{\rm (Scalar~DM)}_{\mu\nu}&=& \partial_\mu S \partial_\nu S-\frac{1}{2}g_{\mu\nu}\partial^\rho S \partial_\rho S+\frac{1}{2}g_{\mu\nu}  m^2_S S^2.
\eea
Here,  $c^G_{X,\chi,S}, c^G_V, c^G_\psi, c^G_H$ are KK graviton couplings which are determined by the overlap between the wave functions of the KK graviton and fields in extra dimensions, see Ref.~\cite{RSbulk} for an example in AdS. $X(\chi,S)$,  $V$ , $H$ and $\psi$ denote the Dark Matter particle, gauge bosons, Higgs and SM matter fields, respectively. 
When the KK graviton mediator connects between dark matter and the SM particles, the DM annihilations are s-wave for scalar and vector dark matters whereas they are p-wave for fermion dark matter. 
Thus, only scalar and vector dark matters lead to observable gamma-ray signatures at present.

The KK graviton couplings to the electroweak gauge bosons are written schematically:
for transverse modes,
\be
 G(c^G_W W_T W_T + c^G_B  B_T B_T) = G(c_{\gamma \gamma} A_TA_T + c_{Z \gamma} Z_T A_T + c_{ZZ} Z_TZ_T + c_{WW} W^+_TW^-_T), 
\ee
with 
\bea
c_{\gamma \gamma} &=& c^G_B \cos^2 \theta_W + c^G_W \sin^2 \theta_W, \nonumber \\
c_{Z \gamma} &=&(c^G_W - c^G_B) \sin(2 \theta_W),  \nonumber \\
c_{ZZ} &=& c^G_W \cos^2 \theta_W + c^G_B \sin^2 \theta_W,  \nonumber \\
c_{WW} &=& 2 c^G_W,
\eea
and for longitudinal modes,
\be
c^G_H G ( m^2_W W^+_LW^-_L + m^2_Z Z_LZ_L ).
\ee
Thus, for the universal gravity couplings to electroweak gauge bosons, $c^G_W=c^G_B=c^G_V$, the $Z\gamma$ coupling vanishes, so there is no DM annihilation into $Z\gamma$ with graviton mediator~\footnote{This result could change if localized kinetic terms~\cite{loc-kin} are introduced, but their effect is naturally suppressed.}. For comparison, non-gravitational interactions of singlet pseudo-scalar or extra gauge boson mediator lead to the DM annihilation into $Z\gamma$ \cite{lpp,lps}, which is an extra source for monochromatic photons\footnote{In the case of box-shaped gamma-ray spectrum\cite{ibarra,jjfan,axionbox}, $Z\gamma$ in the final states comes from the decay of intermediate states that dark matter annihilates into. In this case, $Z\gamma$ channel depends on whether the intermediate state decays into $Z\gamma$ or not.}.

In warped extra-dimensions, there is a hierarchy of couplings of the graviton to Dark matter, Bulk, Matter and Higgs fields, respectively. Indeed, in our setup, one obtains
\bea
& & \textrm{\bf Dark matter: } c^G_X \simeq {\cal O}(1)  \ , \\
& &  \nonumber \\
& & \textrm{\bf Bulk fields : } c^G_V \simeq \frac{1}{\int_{Dark}^{Matter} w(z) \, d z}   \ ,\\
& & \textrm{\bf Matter fields : } c^G_{\psi} = \left(\frac{z_{Matter}}{z_{Dark}}\right)^{\alpha} \ , \\
& &\textrm{\bf Higgs fields: }  c^G_H\simeq  \,\,{\cal O}(1)\,\,/\,\, \left(\frac{z_{Matter}}{z_{Dark}}\right)^{\alpha} \quad {\rm Model\,\,A\,\,/\,\,B}
\label{cs}
\eea
where $\alpha > 1$. In AdS models, the value of $c^G_V$ is
\bea
c^G_V=2 \frac{1-J_0(x_G)}{\log\left(\frac{M_{Pl}}{TeV}\right) \,  x_G^2 \, |J_2(x_G)|} \label{cgamma}
\eea
where $x_G=3.83$ is the first zero of the Bessel function $J_1$, given in the absence of localized kinetic terms. Here we see explicitly the suppression by $(\int w(z) d z)^{-1} = 1/\log(M_P/TeV)\simeq {\cal O}$(0.03).

For simplicity, we have shown the effect of exchanging one KK-mode. In the next sections we will also present results including the effect of the whole KK-tower. This can be done for any metric of the form of Eq.~\ref{metric2} as shown in \cite{SRsa}. See Ref.~\cite{GMDM} for more details.

\subsection{Radion mediator}

The radion of extra dimensions, $r$, couples to the trace of the energy-momentum tensor \cite{csaki,GMDM} as follows,
\bea
{\cal L}_{\rm dilaton}&=&\frac{1}{\sqrt{6}\Lambda}\,r \,T_\mu^\mu \nonumber \\
&=&\frac{1}{\sqrt{6}\Lambda}r \bigg[ T^{\rm DM}+c^r_\psi \Big(-\frac{7}{2}{\bar\psi}i\gamma^\mu D_\mu \psi-\frac{1} {2} D_\mu {\bar\psi}i\gamma^\mu\psi+4m_\psi {\bar\psi}\psi +2\partial^\mu({\bar\psi} i \gamma_\mu\psi)\Big)\nonumber \\
 &&+c^r_H\Big(2 D^\mu H^\dagger D_\mu H-4 V(H)\Big)-\sum_a\frac{\beta_a(g_a)}{2g_a}\,F^a_{\mu\nu} F^{a\mu\nu}
\bigg]
\eea
with
\bea
T^{\rm(Vector\,\,DM)}&=& -  c^r_X m^2_X X_\mu X^\mu, \\
T^{\rm(Fermion\,\,DM)}&=& c^r_\chi\bigg(-3i{\bar\chi}\gamma^\mu\partial_\mu \chi+4m_\chi {\bar\chi}\chi-\frac{5}{2}\partial^\mu({\bar\chi} i \gamma_\mu\chi)\bigg), \\
T^{\rm(Scalar\,\,DM)}&=& c^r_S\Big(-\partial^\mu S \partial_\mu S +2m^2_S S^2\Big).
\eea
Here, the radion couplings are denoted by $c^r_{X,\chi,X}, c^r_\psi, c^r_H$, which are determined by an overlap between the wave functions of the radion and the fields in the extra dimension, similarly as for the KK graviton couplings.

We note that including the linear radion couplings, non-derivative radion interactions to massive scalar and vector particles are fixed by dilatation symmetry \cite{csaki,GMDM} to
\bea
{\cal L}_{\rm non-deriv}&=&-\bigg(\frac{r}{\sqrt{6}\Lambda}-\frac{r^2}{6\Lambda^2}\bigg)\Big(c^r_H m^2_A A_\mu A^\mu+c^r_X m^2_X X_\mu X^\mu\Big) \nonumber \\
&&+2\bigg(\frac{r}{\sqrt{6}\Lambda}
-\frac{r^2}{3\Lambda^2}\bigg)\Big(c^r_H m^2_h h^2+c^r_S m^2_S S^2 \Big ).
\eea

When the radion connects between dark matter and the SM particles, the DM annihilations are s-wave for scalar and vector dark matters whereas they are p-wave for fermion dark matter. 
In the radion case, we note that the interactions to gauge bosons induced by trace anomalies are loop-suppressed, but there exist tree-level couplings to massive gauge bosons after electroweak symmetry breaking. Thus, the radion decay into a photon pair is loop-suppressed, so is the DM annihilation into a photon pair with radion mediator. Therefore, we could not explain the Fermi gamma-ray line with the radion mediator only.

\section{Dark matter annihilations with KK graviton mediator}

In this section, assuming that the KK graviton mediator contributes dominantly to the annihilation processes for dark matter of any spin, we show how the coupling and mass of the KK graviton are constrained by the relic density and the indirect detection experiments.  We also discuss the effect of Higgs portal couplings on the relic density and consider the possibility of explaining the Fermi gamma-ray line in our models. 

First we briefly discuss the model dependence of the dark matter annihilation cross section into a photon pair and the branching fractions of other annihilation channels.
In Model A,  where the SM fermions are localized on the UV brane while gauge bosons propagate in bulk and Higgs boson is localized on the IR brane,  we get 
$c^G_H={\cal O}(1)\gg c^G_V=c^G_g=c^G_W=c^G_B\simeq 0.03\gg c^G_\psi$. In this case, the KK graviton has strong couplings to Higgs boson and longitudinal components of $W,Z$ gauge bosons on the IR brane, while it has suppressed couplings to the transverse components of gauge bosons and the SM fermions.
Then, the branching fraction of the KK graviton decay rate into a photon pair is highly suppressed,
\be
\frac{\Gamma_G(\gamma\gamma)}{\Gamma_G({\rm total})}\simeq\frac{(c^G_B)^2}{8(c^G_g)^2+3(c^G_W)^2+(c^G_B)^2+(c^G_H)^2}\simeq  10^{-3}.
\ee
Thus, assuming the s-channel dominance for DM annihilations, the branching fraction of the DM annihilation into a photon pair to the total cross section is negligibly small in Model A. 

On the other hand, in Model B, all the SM gauge bosons live in bulk, so $c^G_V=c^G_g=c^G_W=c^G_B\simeq 0.03$, whereas the SM fermions and the Higgs doublet are localized at the UV  brane so their couplings to the KK graviton are suppressed as $c^G_\psi, c^G_H\ll c^G_V$.
Then, the KK graviton decays dominantly into the transverse modes of the SM gauge bosons, so there is a definite prediction for the branching fractions for the KK graviton decay. In particular, for universal gauge couplings with $c^G_g=c^G_W=c^G_B$, the branching fraction of the $\gamma\gamma$ decay mode is about 
\be
\frac{\Gamma_G(\gamma\gamma)}{\Gamma_G({\rm total})}\simeq\frac{(c^G_B)^2}{8(c^G_g)^2+3(c^G_W)^2+(c^G_B)^2}=\frac{1}{12}\simeq 0.083.
\ee
Consequently, ignoring $W/Z$ masses and taking only the s-channels with KK graviton mediator for $m_{DM}<m_G$, the ratio of the DM annihilation cross section into $\gamma\gamma$ to the total cross section is roughly the same as the branching fraction of the corresponding KK graviton decay rate and it is about $0.083$ \footnote{As will be shown shortly, W/Z gauge boson masses increase the branching fraction into a photon pair by about $10\%$. We also note that the DM annihilation cross section into a photon pair depends on the velocity of dark matter, so only vector dark matter can explain Fermi-LAT line as will be shown later.}, which could be compatible with the Fermi gamma-ray line \cite{weniger,fermi-lat}. For $m_{\rm DM}>m_G$, dark matter can also annihilate into a pair of KK gravitons, so the branching fractions of annihilation cross sections are changed. In the case of a sizable KK graviton coupling to gluons, the KK graviton can be produced copiously by gluon fusion at the LHC, while the KK graviton to diphoton rate is greater than the Higgs to diphoton rate. Thus, we can constrain the KK graviton coupling to gluons by the Higgs-like boson search at the LHC.

\begin{figure}[t]
\centering
\includegraphics[width=14cm]{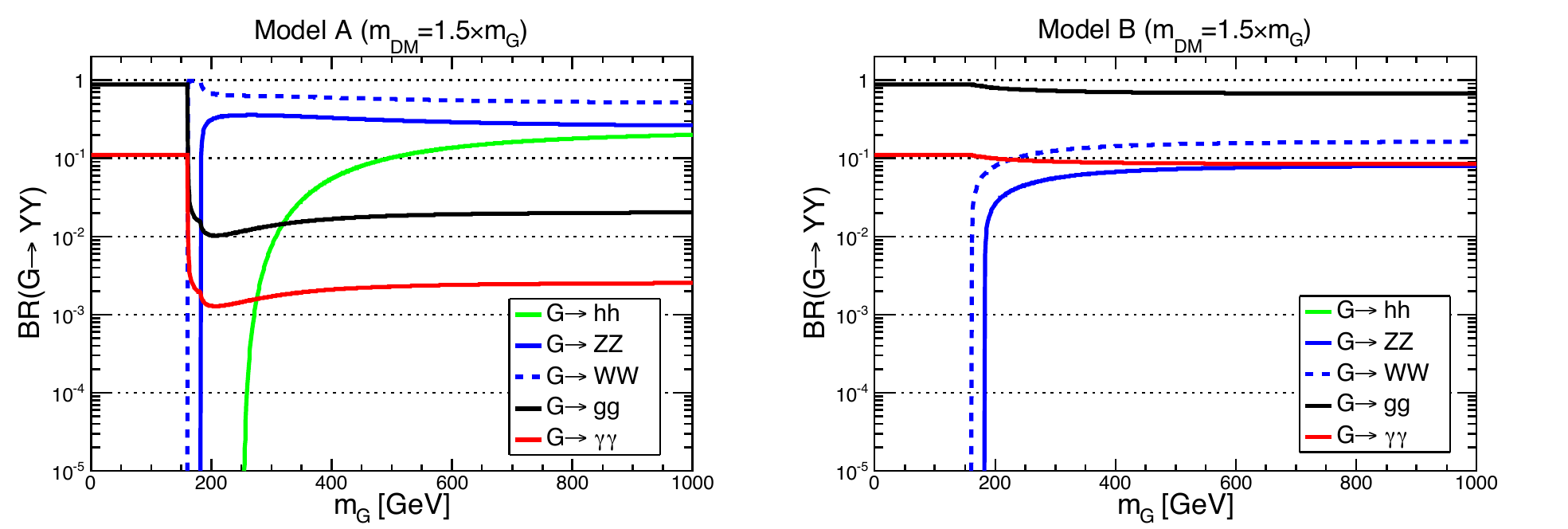} \\
\includegraphics[width=14cm]{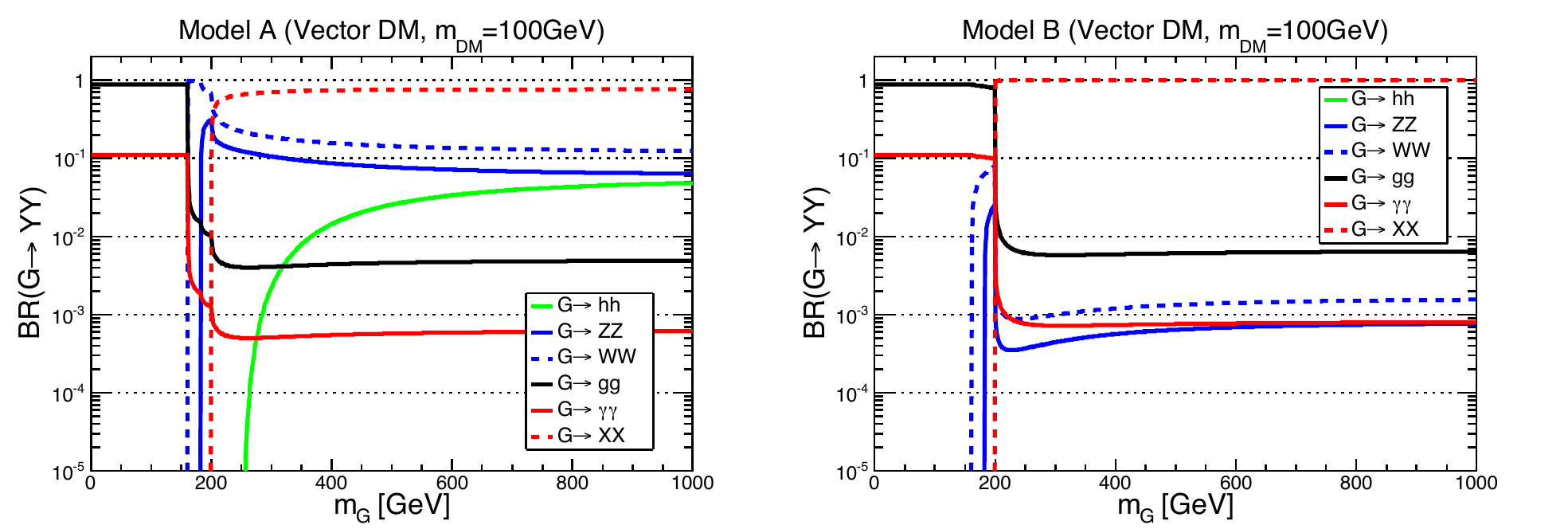}
\caption{Branching fractions of decay rates of the KK graviton for Model A (Left) and B(Right), without (Upper) or with (Lower) the decay mode into a vector dark matter pair for $m_{\rm DM}=100\,{\rm GeV}$.
Green, blue(solid and dashed), black and red lines correspond to branching fractions into $hh,ZZ,WW, gg,\gamma\gamma$, respectively.
We have taken $c_X=1$, $c_V=c_g=c_\gamma=0.03$ in common, and  $c_H=1$ in Model A (Left) and $c_H=0$ in Model B (Right).
}
\label{fig:KKdecay}
\end{figure}

In Fig.~\ref{fig:KKdecay}, the branching fractions of the KK graviton decay rates are shown as a function of KK graviton mass  for Model A and B, respectively. In the upper panel of Fig.~\ref{fig:KKdecay}, we have omitted the decay mode of the KK graviton into a pair of dark matter, assuming that it is kinematically disallowed. In Model A and B, below the $WW$ threshold, the KK graviton decays into a gluon pair with about $90\%$ or a photon pair with about $10\%$. Above the $WW$ threshold, the decay modes into $WW, ZZ$ are dominant in Model A, while they are comparable to the one into a photon pair in Model B. 
On the other hand, if the KK graviton is allowed to decay into a dark matter pair, the extra decay mode becomes dominant and the branching fractions of the other decay modes scale down accordingly. In the lower panel of Fig.~\ref{fig:KKdecay}, the branching fractions of KK graviton decay modes for vector dark matter with $m_{\rm DM}=100\,{\rm GeV}$ are shown. For dark matter with another spin, the extra decay mode into a dark matter pair shows a qualitatively similar behaviour.
We note that the branching fractions of the KK graviton decay also depend on the presence of the decay mode into a top quark pair as shown in Ref.~\cite{GMDM}.

We note that in Model B, if gluons are localized on the UV brane too,
then $c^G_W=c^G_B\simeq 0.03$ and $c^G_\psi, c^G_H, c^G_g\ll c^G_{W,B}$.
Then, for universal electroweak gauge couplings with $c^G_W=c^G_B$, the branching fraction of the $\gamma\gamma$ decay mode becomes about 
\be
\frac{\Gamma_G(\gamma\gamma)}{\Gamma_G({\rm total})}\simeq \frac{(c^G_B)^2}{3(c^G_W)^2+(c^G_B)^2}= 0.25. 
\ee
Thus, in this case, the branching fraction of the DM annihilation into a photon pair would be too large to explain the Fermi gamma-ray line.
However, as will be discussed,  if there are extra annihilation channels coming from radion mediation, it is possible to reduce the branching fraction of the DM annihilation into a photon pair, being consistent with the Fermi gamma-ray line.

Henceforth, we focus on the graviton mediator for dark matter annihilations and denote the KK graviton couplings simply by $c_i$ with $i$ running over the SM particles. 
Before going into the details of each dark matter of a given spin, we summarize the suppression of dark matter annihilation cross sections in Table~\ref{table:summary}.  We note that the $WW/ZZ$ s-channels in Model B are roughly given by those in Model A with $c_H$ being replaced by a volume-suppressed quantity, $c_V$.

\begin{table}[h!] 
\setlength{\tabcolsep}{5pt}
\center
\begin{tabular}{|c|l|c|c|c|c|} 
\hline \hline 
channels & DM mass & X (s=0) & X (s=1/2) & X (s=1)
\\
\hline
s-channel  & $m_{\rm DM}<m_W$ & d-wave & p-wave  & s-wave \\
 \hline
s-channel  & $m_{\rm DM}>m_W$ &  s-wave & p-wave &  s-wave \\
\hline
t-channel & $m_{\rm DM}> m_G$ & s-wave  & s-wave & s-wave \\
\hline \hline
\end{tabular}
\caption{\it Suppression in Dark Matter annihilation cross sections with graviton mediator,  depending on the spin of Dark Matter.}
\label{table:summary} \vspace{-0.35cm}
\end{table}

\subsection{Scalar dark matter}

In Model A, for $c_V\ll c_H$, we get the annihilation cross section of scalar dark matter going into a pair of massive gauge bosons 
\bea
(\sigma v)_{SS\rightarrow Z Z}&\simeq& \frac{3c^2_Sc_H^2}{16\pi \Lambda^4} \frac{m^2_S m^4_Z}{(4m^2_S-m^2_G)^2+\Gamma^2_G m^2_G}\left(1-\frac{4m^2_S}{m^2_G}\right)^2\left(1-\frac{m^2_Z}{m^2_S}\right)^{\frac{1}{2}},  \\
(\sigma v)_{SS\rightarrow W W}&\simeq& \frac{3c^2_Sc_H^2}{8\pi \Lambda^4} \frac{m^2_S m^4_W}{(4m^2_S-m^2_G)^2+\Gamma^2_G m^2_G}\left(1-\frac{4m^2_S}{m^2_G}\right)^2\left(1-\frac{m^2_Z}{m^2_S}\right)^{\frac{1}{2}}
\eea
where
the decay width of the KK graviton is given by
\be
\Gamma_G=\Gamma(hh)+\Gamma(ZZ)+\Gamma(WW)+\Gamma(gg)+\Gamma(SS)
\ee
with
\bea
\Gamma(hh)&=&\frac{c^2_H m^3_G}{960 \pi \Lambda^2} \Big(1-\frac{4m^2_h}{m^2_G}\Big)^\frac{5}{2}, \\
\Gamma(Z Z)&=& \frac{c^2_V m^3_G}{960\pi \Lambda^2}\Big(1- \frac{4m^2_Z}{m^2_G}\Big)^\frac{1}{2}
\Big(13+\frac{56m^2_Z}{m^2_G}+\frac{48m^4_Z}{m^4_G}\Big), \\
\Gamma(W W )&=& \frac{c^2_V m^3_G}{480\pi \Lambda^2}\Big(1- \frac{4m^2_W}{m^2_G}\Big)^\frac{1}{2}
\Big(13+ \frac{56m^2_W}{m^2_G}+\frac{48m^4_W}{m^4_G}\Big), \\
\Gamma(\gamma\gamma)&=& \frac{c^2_\gamma m^3_G}{80\pi\Lambda^2}, \\
\Gamma(gg)&=&  \frac{c^2_g m^3_G}{10\pi\Lambda^2}, \\
\Gamma(SS)&=&  \frac{c^2_S m^3_G}{960 \pi \Lambda^2} \Big(1-\frac{4m^2_S}{m^2_G}\Big)^\frac{5}{2}.
\eea
Moreover, the annihilation cross section into a Higgs pair is d-wave suppressed and it is given by 
\be
(\sigma v)_{SS\rightarrow hh} \simeq v^4 \cdot\frac{ c^2_S c^2_H}{720\pi \Lambda^4}\, \frac{m^6_S}{(4m^2_S-m^2_G)^2+\Gamma^2_G m^2_G} \left(1-\frac{m^2_h}{m^2_S}\right)^\frac{5}{2}.
\ee

In Model B, taking $c_H\ll c_V$, the corresponding annihilation cross sections for scalar dark matter are \cite{GMDM}
\bea
(\sigma v)_{SS\rightarrow Z Z}&\simeq& \frac{3c^2_S c^2_V}{16\pi \Lambda^4} \frac{m^2_S m^4_Z}{(4m^2_S-m^2_G)^2+\Gamma^2_G m^2_G}\left(1-\frac{4m^2_S}{m^2_G}\right)^2\left(1-\frac{m^2_Z}{m^2_S}\right)^{\frac{1}{2}},  \\
(\sigma v)_{SS\rightarrow W W}&\simeq& \frac{3c^2_Sc^2_V}{8\pi \Lambda^4} \frac{m^2_S m^4_W}{(4m^2_S-m^2_G)^2+\Gamma^2_G m^2_G}\left(1-\frac{4m^2_S}{m^2_G}\right)^2\left(1-\frac{m^2_Z}{m^2_S}\right)^{\frac{1}{2}}.
\eea

For both models, we  find that the annihilation cross sections into a photon pair or a gluon pair are always d-wave and are given \cite{GMDM} by
\bea
(\sigma v)_{SS\rightarrow \gamma\gamma}&\simeq&v^4 \cdot  \frac{c^2_S c^2_\gamma }{60\pi\Lambda^4}\frac{m^6_S}{(4m^2_S-m^2_G)^2+\Gamma^2_G m^2_G},\\
(\sigma v)_{SS\rightarrow gg}&\simeq& v^4 \cdot  \frac{2c^2_S c^2_g }{15\pi\Lambda^4}\frac{m^6_S}{(4m^2_S-m^2_G)^2+\Gamma^2_G m^2_G}.
\eea

In the case with heavy dark matter where the contributions of higher KK modes of graviton to the s-channel annihilations become important, we need to perform the following KK sum,
\be
{\cal S}(s)=\frac{1}{\Lambda^2}\sum_{n=1}^\infty  \frac{1}{s-m^2_n+i \,m_n \Gamma_n}. \label{kksum0}
\ee
Here, $\Gamma_n$ denotes the total width of the graviton with KK number $s$ and mass $m_n$ and is given by
\be
\Gamma_n\approx \eta \,m_n \left(\frac{m_n}{\Lambda}\right)^2,\quad\quad \eta=\frac{c^2_H}{240\pi}. 
\ee
The KK graviton masses are determined by the zeros of $J_1(x_n)$ as $m_n=x_n k \Lambda /M_P$, with $x_1=3.83$ and $x_n=\pi(n+1/4)+{\cal O}(n^{-1})$.
For $\eta s\ll \Lambda^2$, we replace the KK graviton propagator with the first KK graviton, $\Lambda^{-2}/(s-m^2_1+i m_1\Gamma_1)$, by the KK sum as follows \cite{GMDM},
\bea
{\cal S}(s)\simeq -\frac{1}{4\Lambda^2\sqrt{s}}\,\frac{x_1}{m_1}\,\frac{J_2(\sigma)}{J_1(\sigma)}
\eea
with $\sigma\simeq (x_1\sqrt{s}/m_1) (1+i\eta s/2\Lambda^2)$.
Henceforth we always take into account the higher KK modes of graviton for our discussion.

Finally, when $m_S>m_G$, there is an extra contribution to the annihilation cross section, due to the t-channel for both models, as follows,
\bea
(\sigma v)_{SS\rightarrow GG} \simeq \frac{4 c_{S}^4 m_{S}^2}{9 \pi \Lambda^4 }
\frac{(1-r_S)^\frac{9}{2}}{r^4_S  (2-r_S)^2}   \label{tch-scalar}
\eea
with $r_S = \left(\frac{m_G}{m_S}\right)^2$.
The t-channel annihilation cross section becomes singular for $r_S\ll 1$, which is a sign of unitarity violation in the case with the KK gravitons in the effective theory. The unitarity bound, $\sigma < 1/s\simeq 1/m^2_S$, implies that $c_S m_G/\Lambda\lesssim (9\pi v)^{1/4} (m_G/m_S)^3$ for a single KK graviton. On the other hand, each higher KK mode contribution is suppressed by $(m_G / m_{KK})^8$ as compared to the first KK graviton contribution, occupying only less than $0.7\%$ of the total t-channel cross section.  Thus, we can ignore the higher KK mode contributions safely.
We note that the t-channel annihilation cross sections for dark matter with another spin show a similar singular behavior.

In Model A and B, the annihilation of scalar dark matter into $WW$ or $ZZ$ is s-wave, unlike the annihilation channels into massless gauge bosons.  But, the corresponding annihilation cross sections are suppressed by $m^4_{Z,W}/m^2_{\rm DM}$, as compared to vector dark matter in the next section.
Thus, for $m_{Z,W}\ll m_{DM}$, the annihilation cross sections of scalar dark matter are much smaller than those of vector dark matter, for a fixed KK graviton coupling. 
If $m_{Z,W}\gtrsim m_{\rm DM}$,  the annihilation cross section of scalar dark matter becomes a sizable s-wave.

\subsection{Fermion dark matter}

In Model A, for $c_H\gg c_V$, the annihilation cross sections for a pair of massive gauge bosons are \cite{GMDM}
\bea
(\sigma v)_{\chi{\bar\chi}\rightarrow ZZ}&\simeq &v^2\cdot \frac{c^2_\chi c^2_H}{144\pi\Lambda^4}\frac{m^6_\chi}{(4m^2_\chi-m^2_G)^2+\Gamma^2_G m^2_G}  \nonumber \\
&&\times\left(1+\frac{3m_Z^2}{m_\chi^2}+\frac{31}{8}\frac{m_Z^4}{m_\chi^4}
-\frac{3m_Z^4}{m_G^2 m_\chi^2}+\frac{6m_Z^4}{m_G^4}
\right)\left(1-\frac{m_Z^2}{m_\chi^2}\right)^\frac{1}{2},~\\
(\sigma v)_{\chi{\bar\chi}\rightarrow WW}&\simeq & v^2\cdot \frac{c^2_\chi c^2_H}{72\pi\Lambda^4}
\frac{ m^6_\chi}{(4m^2_\chi-m^2_G)^2+\Gamma^2_G m^2_G} \nonumber \\
&& \left(1+\frac{3m_W^2}{m_\chi^2}+
 \frac{31}{8}\frac{m_W^4}{m_\chi^4}
-\frac{3m_W^4}{m_G^2 m_\chi^2}+\frac{6m_W^4}{m_G^4}
\right) \left(1-\frac{m_W^2}{m_\chi^2}\right)^\frac{1}{2}
\eea
where the decay rate of the KK graviton is
\be
\Gamma_G=\Gamma(hh)+\Gamma(ZZ)+\Gamma(WW)+\Gamma(gg)+\Gamma(\chi{\bar\chi})
\ee
with
\be
\Gamma(\chi{\bar\chi})= \frac{c_\chi^2 m_G}{160 \pi}
 \left(\frac{m_G}{\Lambda}\right)^2 
 \left(1-\frac{4m^2_\chi}{m^2_G} \right)^\frac{3}{2} \left(1+\frac{8}{3} \frac{m^2_\chi}{m^2_G}\right) .
\ee
Moreover, the annihilation cross section into a Higgs pair is p-wave suppressed and it is given by
\bea
(\sigma v)_{\chi{\bar\chi}\rightarrow hh}\simeq v^2\cdot \frac{c^2_\chi c^2_H }{144\pi\Lambda^4}
\frac{m^6_\chi}{(4m^2_\chi-m^2_G)^2+\Gamma^2_G m^2_G} \left(1-\frac{m^2_h}{m^2_\chi}\right)^\frac{5}{2}.
\eea

On the other hand, in Model B, for $c_H \ll c_V$, the corresponding annihilation cross sections for fermion dark matter are \cite{GMDM}
\bea
(\sigma v)_{\chi{\bar\chi}\rightarrow ZZ} &\simeq&  v^2 \cdot \frac{c_\chi^2 c_V^2}{144\pi \Lambda^4} 
\frac{m_\chi^6}{(m_G^2-4 m_\chi^2)^2+\Gamma_G^2 m_G^2} \,\left(1-\frac{m_Z^2}{m_\chi^2}\right)^\frac{1}{2}\nonumber \\
&&\quad \times\Bigg(12-\frac{9m_Z^2}{m_\chi^2}+\frac{39m_Z^4}{8m_\chi^4}
-\frac{3m_Z^4}{m_G^2 m_\chi^2}+\frac{6m_Z^4}{m_G^4}
\Bigg)  , \\
(\sigma v)_{\chi{\bar\chi}\rightarrow WW} &\simeq&  v^2 \cdot \frac{c_\chi^2 c_V^2}{72\pi \Lambda^4} 
\frac{m_\chi^6}{(m_G^2-4 m_\chi^2)^2+\Gamma_G^2 m_G^2} \, \left(1-\frac{m_W^2}{m_\chi^2}\right)^\frac{1}{2}\nonumber \\
&&\quad\Bigg(12-\frac{9m_W^2}{m_\chi^2}+\frac{39m_W^4}{8m_\chi^4}
-\frac{3m_W^4}{m_G^2 m_\chi^2}+\frac{6m_W^4}{m_G^4}
\Bigg) 
\eea

For both models, we obtain the annihilation cross sections into a pair of massless gauge bosons \cite{GMDM} as
\bea
(\sigma v)_{\chi{\bar\chi}\rightarrow \gamma\gamma}&\simeq& v^2\cdot \frac{c^2_\chi c^2_\gamma}{12\pi\Lambda^4}\frac{m^6_\chi}{(4m^2_\chi-m^2_G)^2+\Gamma^2_G m^2_G},\\
(\sigma v)_{\chi{\bar\chi}\rightarrow gg}&\simeq&  v^2\cdot\frac{2c^2_\chi c^2_g}{3\pi\Lambda^4}\frac{m^6_\chi}{(4m^2_\chi-m^2_G)^2+\Gamma^2_G m^2_G}.
\eea
On the other hand, when $m_{\chi}>m_G$, there is an extra contribution to the annihilation cross section, due to the t-channel for both models, as follows,
\bea
(\sigma v)_{\chi \bar\chi \rightarrow GG} &\simeq& \frac{c_{\chi}^4 m_{\chi}^2}{16 \pi \Lambda^4 }
\frac{(1-r_\chi)^\frac{7}{2}}{r^4_\chi (2-r_\chi)^2}  \label{tch-fermion}
\eea
with $r_\chi = \left(\frac{m_\chi}{m_S}\right)^2$.

\begin{figure}[t]
\centering
\includegraphics[width=7cm]{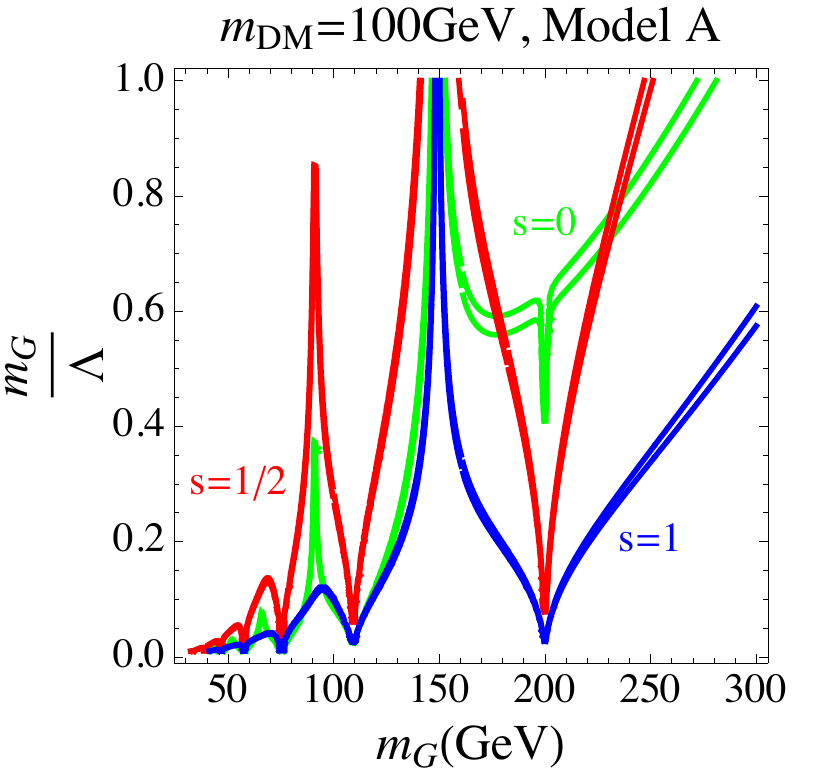}
\includegraphics[width=7cm]{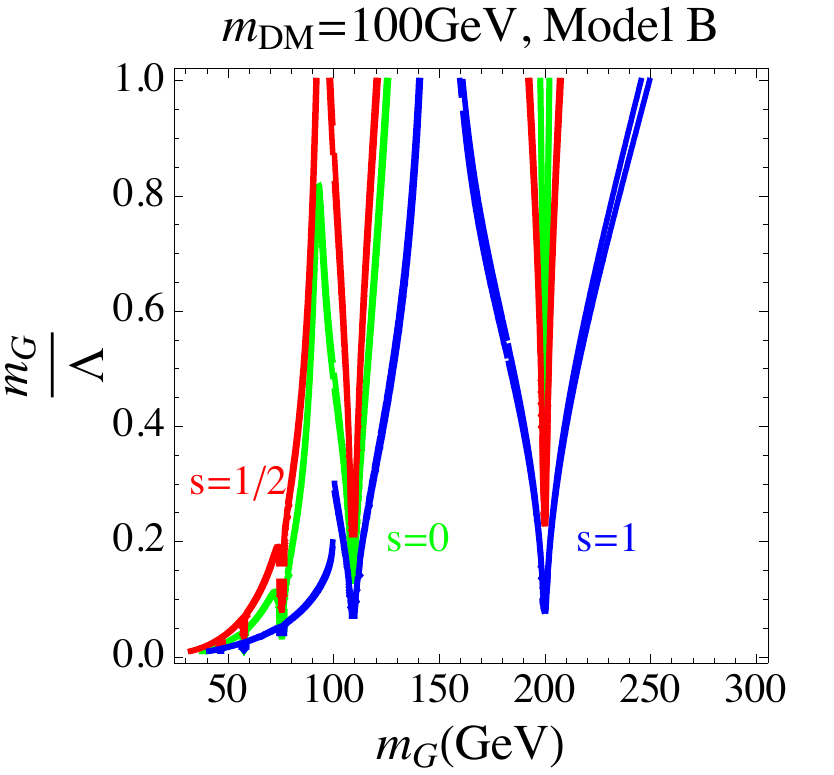}
\caption{Parameter space of the effective DM coupling, $m_G/\Lambda$, vs $m_G$ for scalar(Green), fermion(Red) and vector(Blue) dark matters, satisfying the relic density condition. 
We have taken $c_X=1$, $c_V=c_g=c_\gamma=0.03$ in common, and  $c_H=1$ on left (Model A) and $c_H=0$ on right (Model B). Here, we note that spikes appearing for $m_G<2 m_{\rm DM}$ correspond to resonances due to the higher KK modes of graviton.
}
\label{fig:relicSF}
\end{figure}

Consequently, in Model A and B, for $m_G> m_{\rm DM}$, the annihilation cross sections of fermion dark matter are p-wave suppressed. On the other hand, for  $m_G< m_{\rm DM}$, the t-channel annihilation into a pair of the KK gravitons is s-wave and becomes dominant in determine the relic density. 
As shown in Fig.~\ref{fig:relicSF}, for $m_G< m_{\rm DM}$, 
scalar or fermion dark matter can account for the relic density for small values of the effective KK graviton couplings. On the other hand, for $m_G> m_{\rm DM}$, it is the s-channel annihilations that determine the relic density, so the necessary KK graviton couplings should be larger than in the case with $m_G< m_{\rm DM}$.
If radion mediation is included, of course, the necessary effective KK graviton coupling can be smaller for the relic density in both cases \cite{GMDM}.

\subsection{Vector dark matter}

In Model A, for $c_V \ll c_H$, the annihilation cross sections of vector dark matter going into a pair of massive gauge bosons are
\bea
(\sigma v)_{XX\rightarrow ZZ} &\simeq&  \frac{2 c_{X}^2 c_H^2}{27\pi \Lambda^4} 
\frac{m_{X}^6}{(m_G^2-4 m_{X}^2)^2+\Gamma_G^2 m_G^2}  \nonumber \\
&&\times \left(1+\frac{3m_Z^2}{m_{X}^2}+\frac{115}{32}\frac{m_Z^4}{m_{X}^4}
-\frac{3}{4}\frac{m_Z^4}{m_G^2 m_{X}^2}+\frac{3}{2}\frac{m_Z^4}{m_G^4}
\right) \left(1-\frac{m_Z^2}{m_{X}^2}\right)^\frac{1}{2} , \\
(\sigma v)_{XX\rightarrow WW} &\simeq&  \frac{4 c_{X}^2 c_H^2}{27\pi \Lambda^4} 
\frac{m_{X}^6}{(m_G^2-4 m_{X}^2)^2+\Gamma_G^2 m_G^2}  \nonumber \\
&&\times \left(1+\frac{3m_W^2}{m_{X}^2}+\frac{115}{32}\frac{m_W^4}{m_{X}^4}
-\frac{3}{4}\frac{m_W^4}{m_G^2 m_{X}^2}+\frac{3}{2}\frac{m_W^4}{m_G^4}
\right)
 \left(1-\frac{m_W^2}{m_{X}^2}\right)^\frac{1}{2} 
\eea 
where the decay width of the KK graviton is
\be
\Gamma_G=\Gamma(hh)+\Gamma(ZZ)+\Gamma(WW)+\Gamma(gg)+\Gamma(XX)
\ee
with
\bea
\Gamma(XX)&=& \frac{c^2_X m^3_G}{960\pi \Lambda^2}\Big(1- \frac{4m^2_X}{m^2_G}\Big)^\frac{1}{2}
\Big(13+\frac{56m^2_X}{m^2_G}+\frac{48m^4_X}{m^4_G}\Big).
\eea
Moreover, the annihilation cross section into a Higgs pair is s-wave and is given by
\bea
(\sigma v)_{XX\rightarrow hh}\simeq \frac{2c^2_X c^2_H}{27\pi \Lambda^4}
\frac{m^6_X}{(4m^2_X-m^2_G)^2+\Gamma^2_G m^2_G}\left(1-\frac{m^2_h}{m^2_X}\right)^\frac{5}{2}.
\eea

On the other hand,  in Model B, $c_H \ll c_V$, the corresponding annihilation cross sections for vector dark matter  are \cite{GMDM}
\bea
(\sigma v)_{XX\rightarrow ZZ} &\simeq&  \frac{2 c_{X}^2 c_V^2}{27\pi \Lambda^4} 
\frac{m_{X}^6}{(m_G^2-4 m_{X}^2)^2+\Gamma_G^2 m_G^2}\, \left(1-\frac{m_Z^2}{m_{X}^2}\right)^\frac{1}{2}\nonumber \\
&&\quad\times\Bigg(12-\frac{9m_Z^2}{m_{X}^2}+\frac{147}{32}\frac {m_Z^4}{m_{X}^4} 
-\frac{3}{4}\frac{m_Z^4}{m_G^2 m_{X}^2}+\frac{3}{2}\frac{m_Z^4}{m_G^4}
\Bigg)  , \\
(\sigma v)_{XX\rightarrow WW} &\simeq&  \frac{4 c_{X}^2 c_V^2}{27\pi \Lambda^4} 
\frac{m_{X}^6}{(m_G^2-4 m_{X}^2)^2+\Gamma_G^2 m_G^2}\, \left(1-\frac{m_W^2}{m_{X}^2}\right)^\frac{1}{2}  \nonumber \\
&&\times \Bigg(12-\frac{9m_W^2}{m_{X}^2}+\frac{147}{32}\frac {m_W^4}{m_{X}^4} 
-\frac{3}{4}\frac{m_W^4}{m_G^2 m_{X}^2}+\frac{3}{2}\frac{m_W^4}{m_G^4}.
\Bigg) 
\eea

For both models, we also obtain the annihilation cross sections into a pair of massless gauge bosons \cite{GMDM} as
\bea
(\sigma v)_{XX\rightarrow \gamma\gamma}&=&\frac{8c^2_X c^2_\gamma}{9\pi\Lambda^4}\frac{m^6_X}{(4m^2_X-m^2_G)^2+\Gamma^2_G m^2_G}, \\
(\sigma v)_{XX\rightarrow gg}&=&\frac{64c^2_X c^2_g}{9\pi\Lambda^4}\frac{m^6_X}{(4m^2_X-m^2_G)^2+\Gamma^2_G m^2_G}.
\eea
On the other hand, when $m_X>m_G$, there is an extra contribution to the annihilation cross section, due to the t-channel in both models, as follows,
\bea
(\sigma v)_{X X  \rightarrow GG} &\simeq&
\frac{c_{X}^4 m_{X}^2}{324 \pi \Lambda^4 } 
\frac{\sqrt{1-r_X}}{r^4_X  (2-r_X)^2} \,
\bigg(176+192 r_X+1404 r^2_X-3108 r^3_X \nonumber \\
&&+1105 r^4_X+362 r^5_X+34 r^6_X \bigg)
\eea
with $r_X = \left(\frac{m_G}{m_X}\right)^2$.
Therefore, the total annihilation cross sections for vector dark matter are s-wave.
Furthermore, the partial annihilation cross section into a photon pair is sizable due to the universal gravity couplings to gauge bosons, $c_V=c_\gamma=c_g$.

In Fig.~\ref{fig:relicSF}, we have shown the parameter space of the KK graviton coupling vs the KK graviton mass for a fixed mass of vector dark mass, in comparison to the cases of scalar and fermion dark matters. We note that the annihilation cross section of vector dark matter is s-wave independent of dark matter mass, as compared to the cases of scalar and fermion dark matters.
There, vector dark matter is strongly subject to the gamma-ray detection experiments at present, as will be shown later.

\subsection{Impact of Higgs portal couplings}

\begin{figure}[t]
\centering
\includegraphics[width=7.1cm]{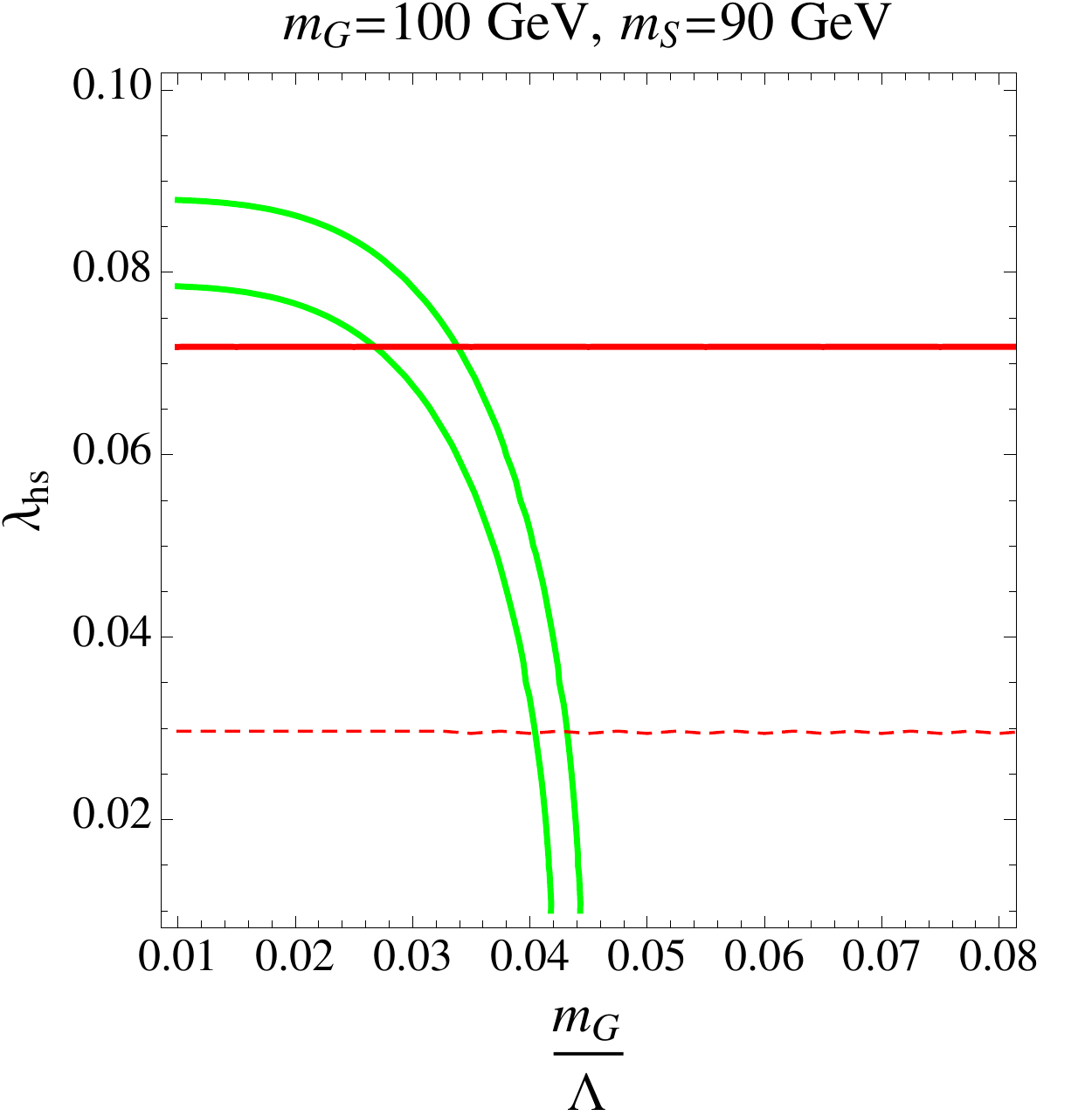}\\
 \caption{The KK graviton coupling versus Higgs portal coupling for scalar dark matter.   The Planck $5\sigma$ band for the relic density is bounded by green lines while the spin-independent DM-nucleon scattering cross section, $\sigma_{S-N}=10^{-9}\,{\rm pb}$, for the minimum and maximum values of Higgs-nucleon  couplings, are shown in red solid and dashed lines, respectively \cite{mambrini}.  
}
\label{fig:Higgsportal}
\end{figure}

In this section, we discuss the effect of a Higgs portal coupling on our analysis of the DM annihilation cross sections and the direct detection of dark matter.  In the case of Model A, where the Higgs doublet and dark matter are localized on the same IR brane, a renormalizable Higgs portal coupling to the scalar dark matter, ${\cal L}\supset -\lambda_{hS} H^\dagger H S^2/4$, is allowed by any symmetry of our model setup.  Non-renormalizable Higgs portal couplings to singlet fermion or vector dark matter can be written too but they depend on a UV completion. So, we focus on the case of scalar dark matter. In the case of Model B, there is no Higgs portal coupling at tree level, because the Higgs doublet and dark matter are localized on different branes. Instead, an effective quadratic Higgs coupling to dark matter is generated by the bulk graviton mediators, but there is no linear Higgs coupling to dark matter due to the fact that the KK graviton couples to the energy-momentum tensor.

There are three main effects of the Higgs portal coupling to scalar dark matter. 
First, it contributes to the DM annihilation cross section with extra s-channels of Higgs mediator. Second, when scalar dark matter is lighter than half the Higgs mass, the Higgs decay rates are affected by the invisible decay of Higgs into a pair of dark matter.  From the search for the invisible decay of Higgs at the LHC, the branching fraction of Higgs invisible decay is constrained to ${\rm Br}(h\rightarrow SS)<0.65$ at $95\%$ C.L. \cite{Hinv}. For such a light dark matter below the WW threshold, the s-channel DM annihilations with KK graviton mediator are velocity-suppressed, so the extra annihilation channels with Higgs mediator can be dominant, unless the t-channel annihilation into a pair of the KK graviton is open.
Third, the Higgs portal coupling could lead to a sizable spin-independent cross section between dark matter and nucleons so it is strongly constrained by the direct detection experiments \cite{xenon100,lux}, in particular, below the WW threshold.

In Fig.~\ref{fig:Higgsportal}, focusing on the region above the WW threshold, we depict the parameter space for the KK graviton coupling, $m_G/\Lambda$, vs the Higgs portal coupling $\lambda_{hS}$, satisfying the relic density with direct detection constraints. Whether the t-channel exchange is open or closed (as in the choice of masses in Fig.~\ref{fig:Higgsportal}) the qualitative behaviour is the same. We note that the correct relic density can be obtained from a combination of the KK graviton coupling and the Higgs portal coupling, being consistent with the direct detection experiments.

\subsection{The Fermi-LAT line}

\begin{figure}[t]
\centering
\includegraphics[width=8cm]{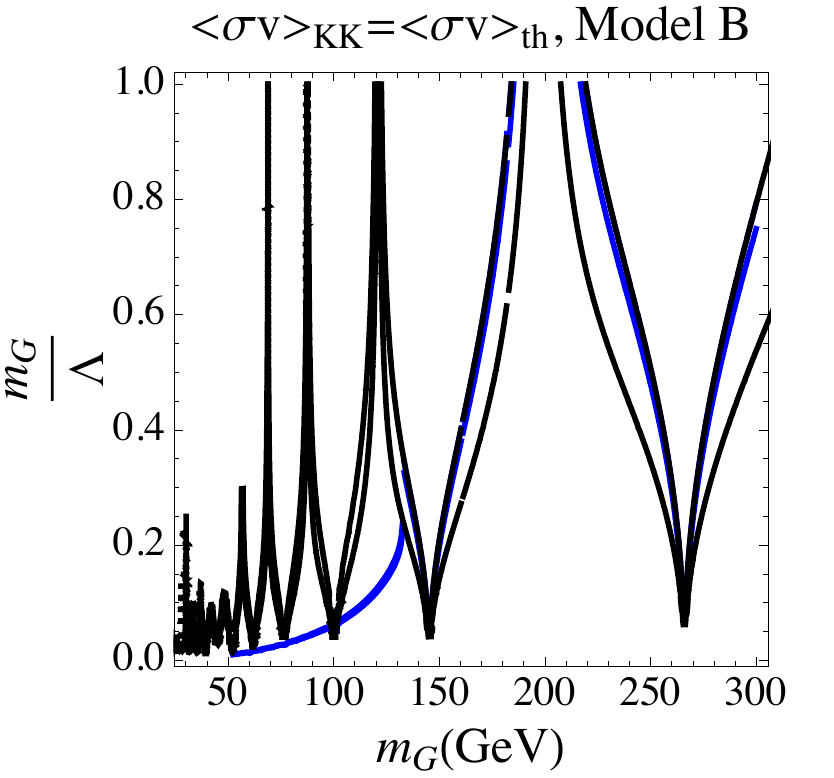}
\caption{Parameter space of the effective DM coupling $m_G/\Lambda$ vs $m_G$ for vector dark matter, satisfying the relic density condition (Blue) and the Fermi gamma-ray line at zero temperature (Black). 
We have set $m_X$=133 GeV for Fermi gamma-ray line with $\langle\sigma v\rangle_{\gamma\gamma}=(0.67-3.16)\times 10^{-27}{\rm cm}^3/{\rm s}$ \cite{weniger} (within $1\sigma$ range for NFW and Einasto dark matter profiles)  and imposed the relic density within Planck $5\sigma$ band. We have taken $c_X=1$, $c_H=c_f=0$ and $c_V=c_\gamma=c_g=0.03$.  
}
\label{fig:Fermi-gray}
\end{figure}

We are in a position to discuss briefly the possibility of obtaining the Fermi gamma-ray line in the models. Most dark matter models explaining the Fermi gamma-ray line require new charged and/or neutral mediators \cite{models} and invoke a strong coupling or a resonance in order to explain a large annihilation cross section into a photon pair for Fermi gamma-ray line at around $130\,{\rm GeV}$. 
While couplings of dark matter or mediator field to a photon pair depend on unknown new charged particles in most cases in the literature, in our gravity-mediated dark matter models,  the couplings of the KK graviton mediator to a photon pair depend on the bulk profiles of KK gravitons and a photon.

For scalar dark matter, the partial annihilation cross section into a photon pair is d-wave suppressed, so scalar dark matter does not give rise to a sizable branching fraction of monochromatic photons
for the Fermi gamma-ray line.  Likewise, for fermion dark matter, 
the annihilation cross section into a photon pair is p-wave suppressed at present, so fermion dark matter with graviton mediator is not relevant for the Fermi gamma-ray line either. For both scalar and fermion dark matter, extra annihilation channels coming from radion mediation \cite{GMDM} could reduce the branching fraction of monochromatic photons further, because the photon channel with radion mediation is loop-suppressed.

Unlike the previous cases, vector dark matter can accommodate the Fermi gamma-ray line in some cases.
First, in Model A, the branching fraction for the DM annihilation into a photon pair is very small, due to large annihilation cross section into a pair of gauge bosons. Thus, we could not obtain the Fermi gamma-ray line in this case and there is no gamma-ray constraint for monochromatic photons.
On the other hand, in Model B, when the t-channel annihilation into  a pair of KK gravitons is forbidden kinematically for $m_G > m_X$, the branching fraction of the DM annihilation cross section into a photon pair can be sizable as the following,
\be
\frac{(\sigma v)_{\gamma\gamma}}{(\sigma v)_{\rm tot}}\simeq \frac{4}{9}\left\{4+\frac{1}{9}\Big(4-3\frac{m^2_Z}{m^2_X}+\frac{49}{32}\frac{m^4_Z}{m^4_X}-\frac{1}{4}\frac{m^4_Z}{m^2_G m^2_X}+\frac{1}{2}\frac{m^4_Z}{m^4_G}\Big)\sqrt{1-\frac{m^2_Z}{m^2_X}}+2(m_Z\rightarrow m_W)\right\}^{-1}.
\ee
Therefore, we find that taking vector dark matter mass to be $m_X\simeq 133\,{\rm GeV}$ for Fermi gamma-ray line \cite{fermi-lat}, the $W/Z$ mass dependence of the annihilation cross sections makes the branching fraction into a photon pair larger, with $(\sigma v)_{\gamma\gamma}/(\sigma v)_{\rm KK}\simeq 0.093(0.36)$ in Model B with $c_g=0.03(c_g\simeq 0)$. 
In Fig.~\ref{fig:Fermi-gray}, we depict the parameter space accounting for the relic density with the KK graviton explaining the total relic density, $\langle\sigma v\rangle_{\rm KK}=\langle\sigma v\rangle_{\rm th}$, within $5\sigma$ of the Planck data, $\Omega h^2=0.1199\pm 0.0027$ \cite{planck}, and the annihilation cross section into a photon pair necessary for the Fermi gamma-ray line \cite{weniger,fermi-lat}.
To conclude, we find that for vector dark matter in Model B with all gauge bosons in bulk, both the relic density and the Fermi gamma-ray line are satisfied with the KK graviton mediator only, with a sizable effective KK graviton coupling, away from the resonance.

\section{Astrophysical constraints}

\begin{figure}[t]
\centering
\includegraphics[width=7cm]{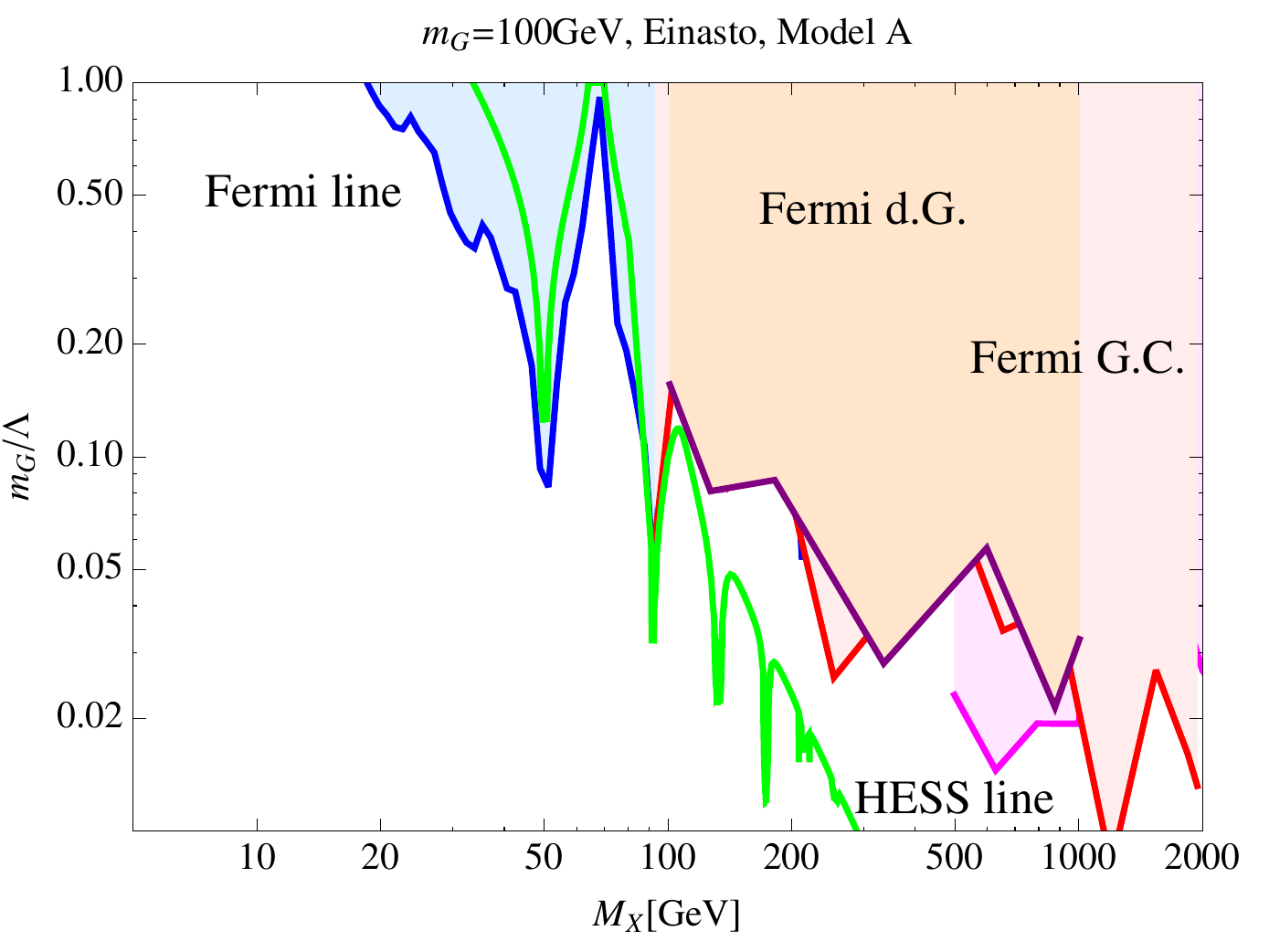}
\includegraphics[width=7cm]{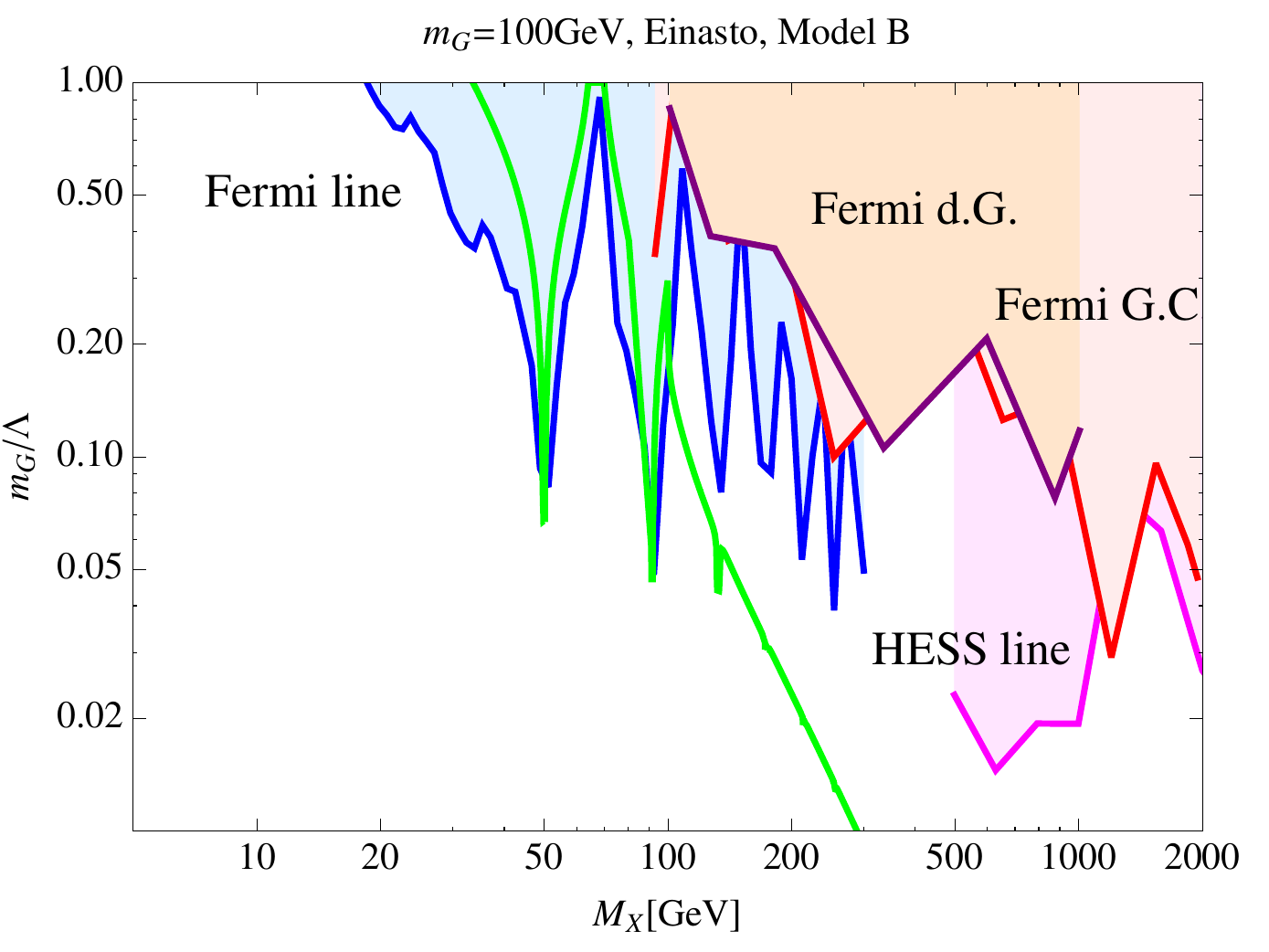}
\caption{Astrophysical bounds on the parameter space, $M_X$ vs $m_G/\Lambda$, for vector dark matter.
We have imposed the bounds from Fermi-LAT and HESS line searches, Fermi-LAT dwarf galaxies (d.G.) and Fermi-LAT galactic center (G.C.)  on the annihilation cross section for Einasto dark matter profile. Green dashed lines show the Planck $5\sigma$ band for the relic density.
We have taken $c_X=1$, $c_V=c_g=c_\gamma=0.03$ in common, and  $c_H=1$ on left (Model A) and $c_H=0$ on right (Model B).
}
\label{fig:astrobounds}
\end{figure}

\begin{figure}[t]
\centering
\includegraphics[width=7cm]{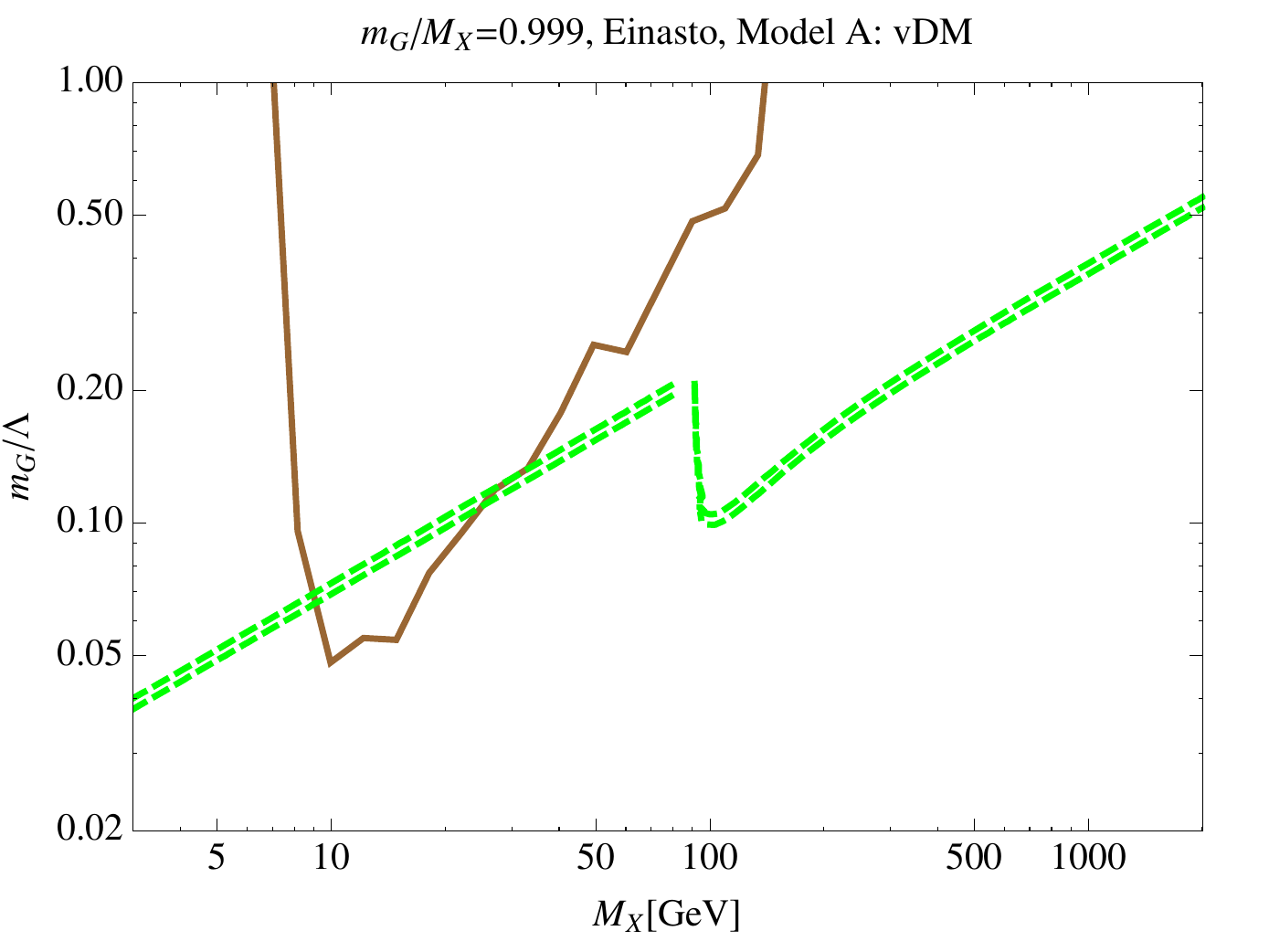}
\includegraphics[width=7cm]{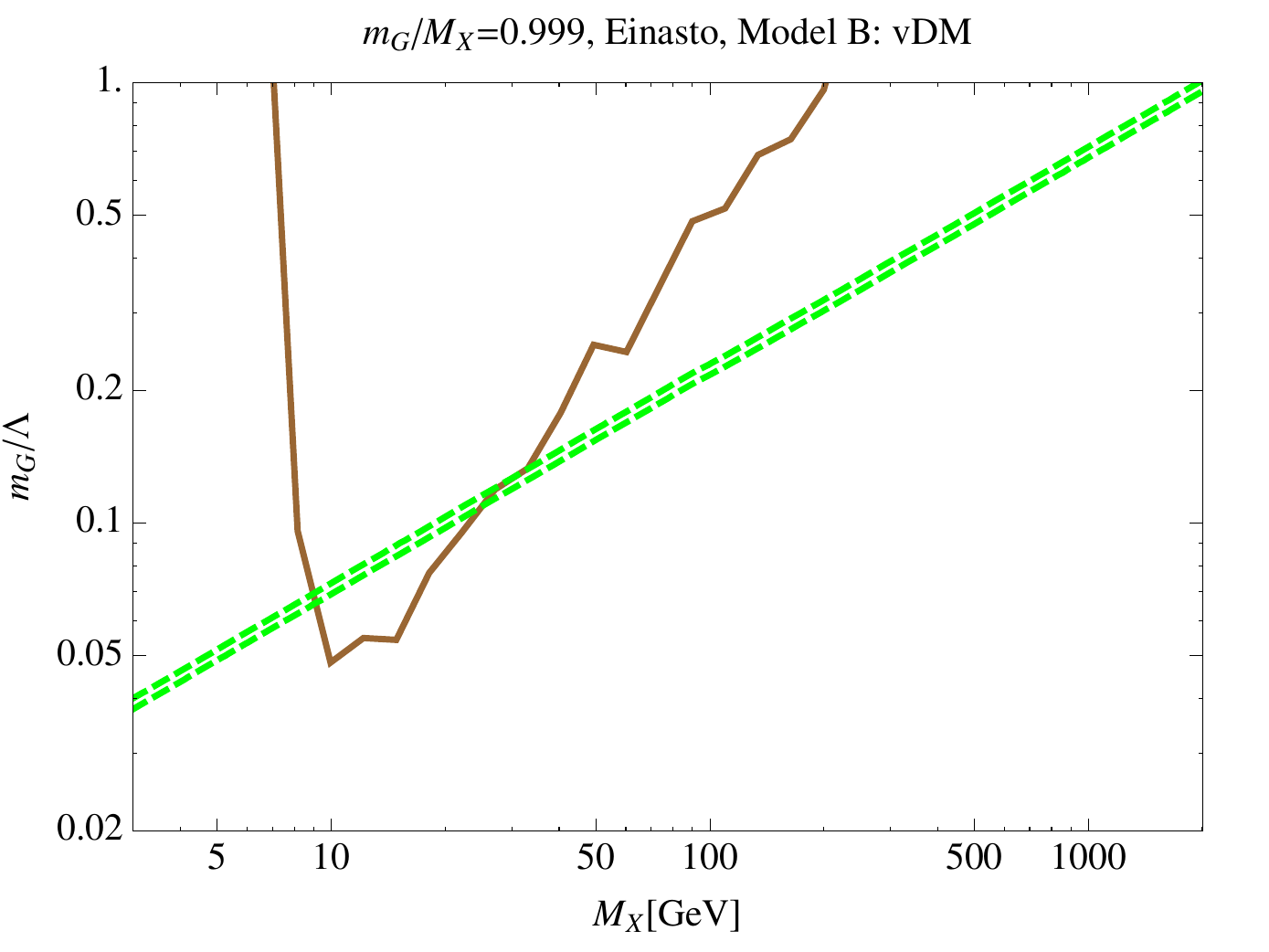}
\caption{Bounds from narrow gamma-ray boxes for vector dark matter.
We have imposed the bounds from Fermi-LAT galactic center (R16) on the annihilation cross section, $\langle\sigma v\rangle_{XX\rightarrow GG}\times {\rm Br}(G\rightarrow \gamma\gamma)$, for Einasto dark matter profile \cite{axionbox}. Green dashed lines show the Planck $5\sigma$ band for the relic density.
We have taken $m_G/M_X=0.999$, $c_X=1$, $c_V=c_g=c_\gamma=0.03$ in common, and  $c_H=1$ on left (Model A) and $c_H=0$ on right (Model B).
}
\label{fig:box1}
\end{figure}

We consider astrophysical constraints on dark matter models with KK graviton mediator.
Typical gamma-ray features are composed of monochromatic photons, continuum photons and gamma-ray boxes. We impose gamma-ray constraints from Fermi-LAT and HESS on the model parameters.
In Fig.~\ref{fig:astrobounds}, we show the bounds from monochromatic and continuum photons on the $s$-channels of the models.  In Fig.~\ref{fig:box1}, the gamma-ray constraints are given for the $t/u$-channels, which are responsible for the gamma-ray boxes. In both cases, we impose the relic density condition for the necessary total thermal cross section.

\subsection{Monochromatic photons}

In models with KK graviton mediator, dark matter annihilation into a photon pair leads to monochromatic photons at the energy of DM mass, so it can be constrained by gamma line searches at Fermi-LAT \cite{fermi-lat} and HESS \cite{hess2013}. 
From Fig.~\ref{fig:astrobounds}, for vector dark matter, it is shown that most of the parameter space below the $WW$ threshold explaining the relic density condition can be ruled out or in tension with the Fermi-LAT bounds, due to a sizable branching fraction of dark matter annihilation into a photon pair. 
For $M_X>m_G$, the Fermi-LAT bounds can be weakened due to the t-channel dominance.  Nonetheless, a certain range of small KK graviton masses satisfying $M_X >m_G$ could be also ruled out by the gamma-ray box constraints as will be discussed later. On the other hand, there is no gamma-ray bound on scalar or fermion dark matter, as the annihilation cross sections into a photon pair are d-wave or p-wave suppressed.

\subsection{Continuum photons}

When DM annihilates into W/Z gauge bosons or SM fermions, continuum photons can be generated from the secondary processes and they can be bounded by Fermi-LAT dwarf galaxies \cite{dwarfgalaxy}, the gamma-ray from Fermi-LAT galactic center \cite{gc} and PAMLA anti-proton data \cite{antiproton,hambye}. We consider the first two bounds as they are more stringent than or as strong as the last one\footnote{For vector dark matter below WW threshold in Model A and for $m_{DM}< m_G$ in Model B, the branching fraction of the $gg$ channel is about $90\%$ the total annihilation cross section at present. The anti-proton bound on $\langle\sigma v\rangle_{gg}$ is about $10^{-26}{\rm cm}^3/{\rm s}$ for MED or MIN propagation parameters \cite{hambye}, which is not as strong as the Fermi-LAT bound.}. In Fig.~\ref{fig:astrobounds}, for vector dark matter, we imposed the above bounds on the continuum photons in addition to Fermi-LAT and HESS monochromatic photons, as denoted in the figures, and found that the current HESS or continuum bounds are the strongest for heavy dark matter with $m_X\gtrsim m_G$, but they are compatible with the relic density condition.  
We note that for scalar dark matter, the $WW/ZZ$ channels, if kinematically open, are s-wave and dominant for $m_X< m_G$, so there is a similar strong bound from continuum photons, as for vector dark matter.  On the other hand, for fermion dark matter, there is no bound from continuum photons.

\subsection{Gamma-ray boxes}

For the t-channel annihilation of dark matter into a pair of KK gravitons, the consequent decay of each KK graviton into a photon pair gives rise to a box-shaped gamma-ray spectrum \cite{ibarra,jjfan,axionbox,gbox,ibarra2}.  Once the t-channel is open,  it becomes dominant for determining the relic density.  Thus, when the decay branching fraction of the KK graviton into a photon pair is sizable, it can be strongly constrained by Fermi-LAT and HESS gamma-ray constraints \cite{ibarra,axionbox}.   

In Fig.~\ref{fig:box1}, we first consider the bounds on narrow gamma-ray boxes and show the Fermi-LAT bounds (R16 for Einasto dark matter profile \cite{fermi-lat}) for vector dark matter models. Taking $m_G/M_X=0.999$ as an example, we find that vector dark matter masses of $9-30\,{\rm GeV}$ have been ruled out by the Fermi-LAT bounds.
On the other hand, in the case of scalar or fermion dark matter, there is no bound on narrow gamma-ray boxes, because the corresponding t-channel annihilation cross sections, (\ref{tch-scalar}) and (\ref{tch-fermion}), are highly suppressed for almost degenerate dark matter and KK graviton masses.

\begin{figure}[t]
\centering
\includegraphics[width=5cm]{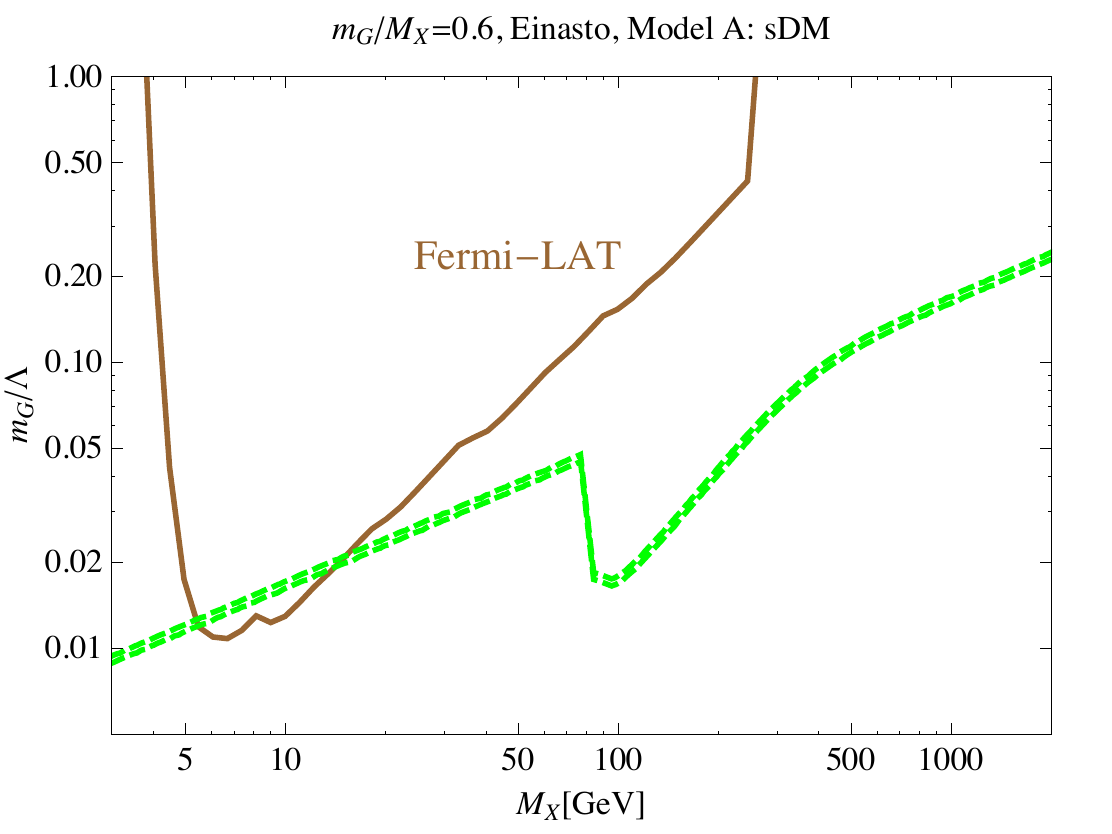}
\includegraphics[width=5cm]{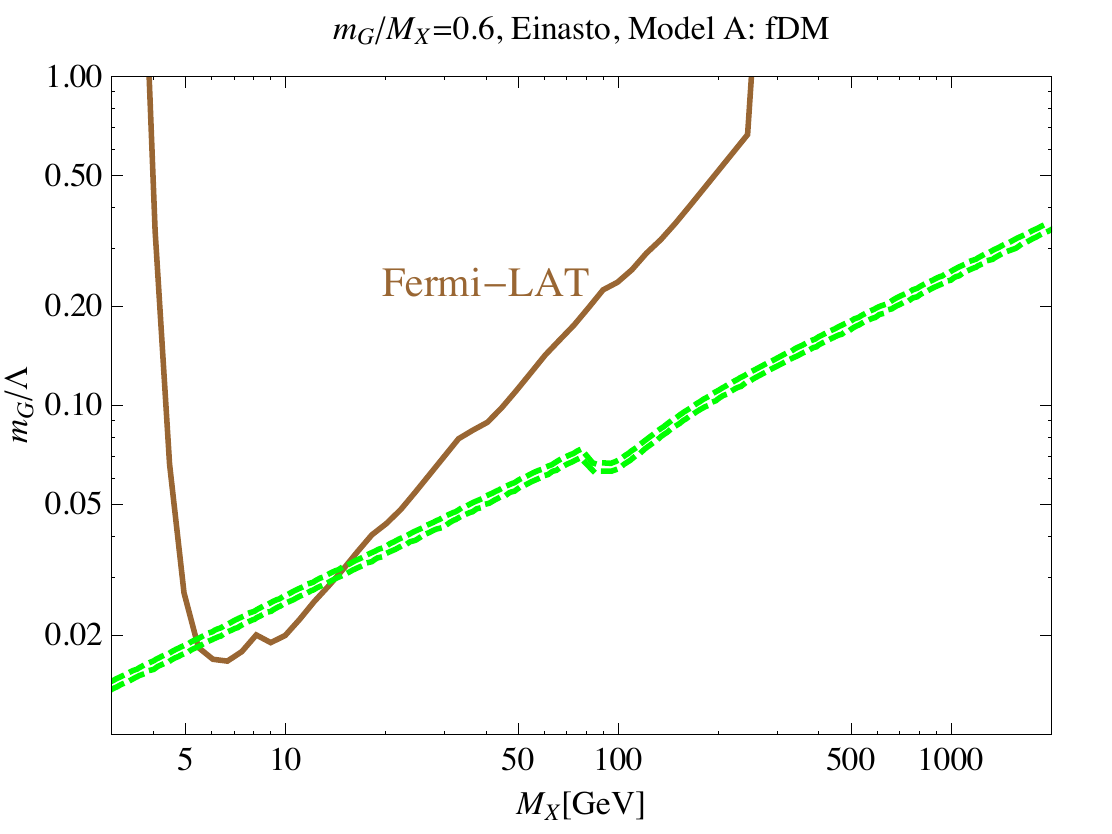}
\includegraphics[width=5cm]{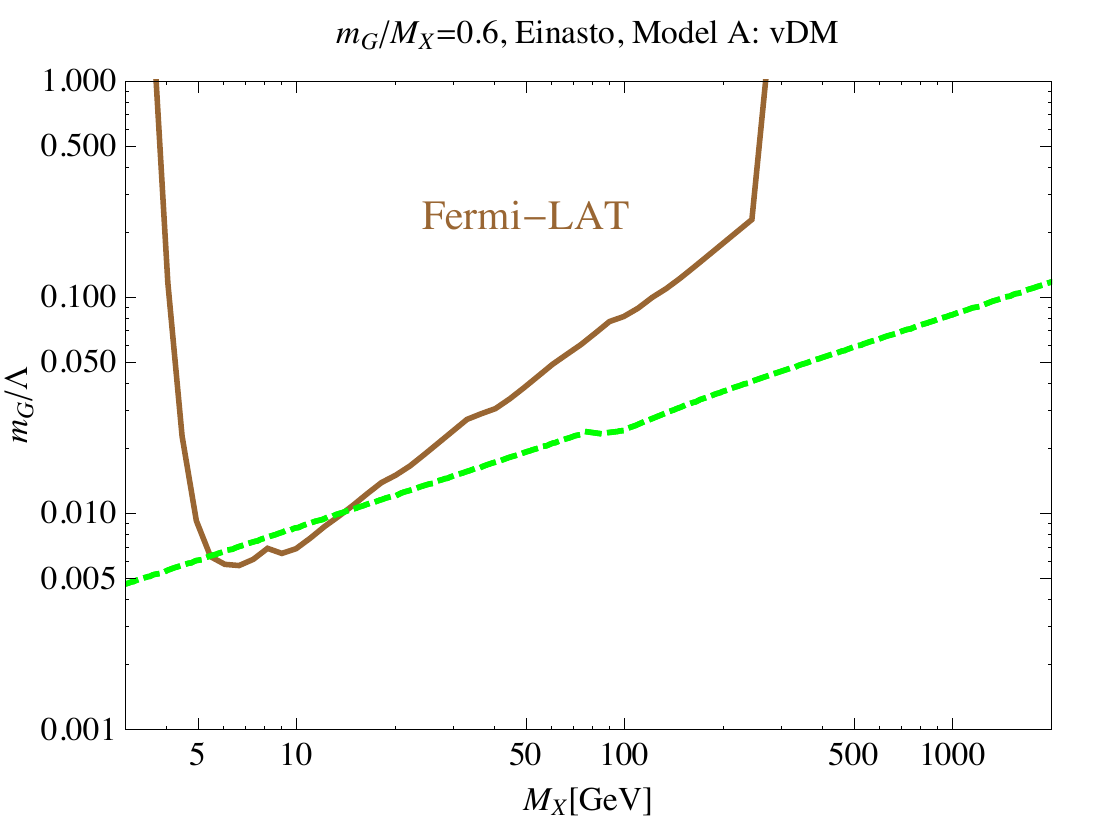}
\\
\includegraphics[width=5cm]{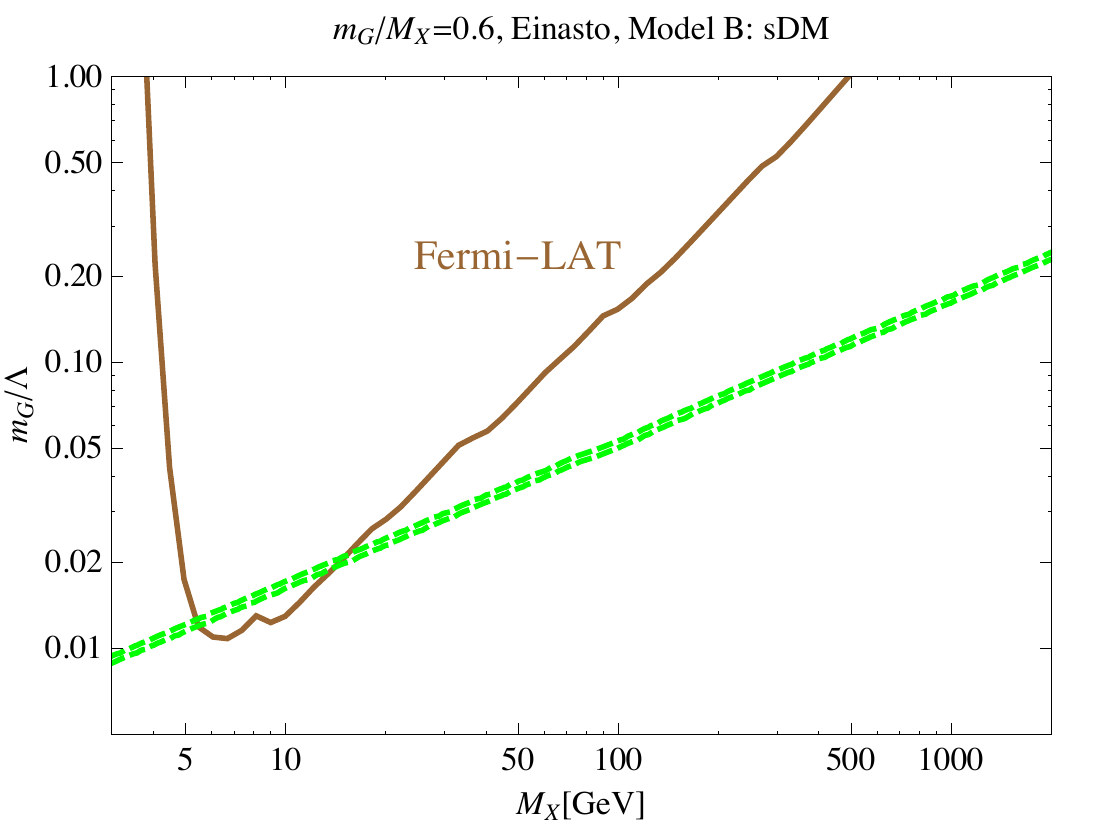}
\includegraphics[width=5cm]{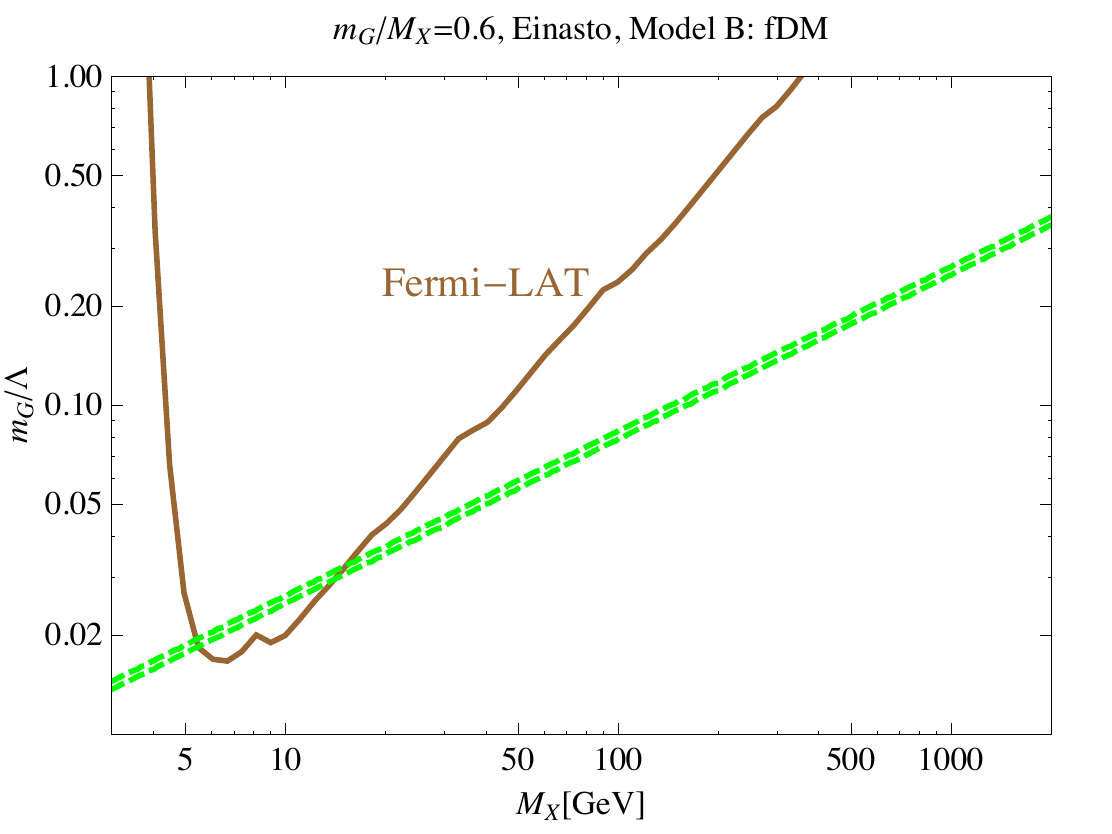}
\includegraphics[width=5cm]{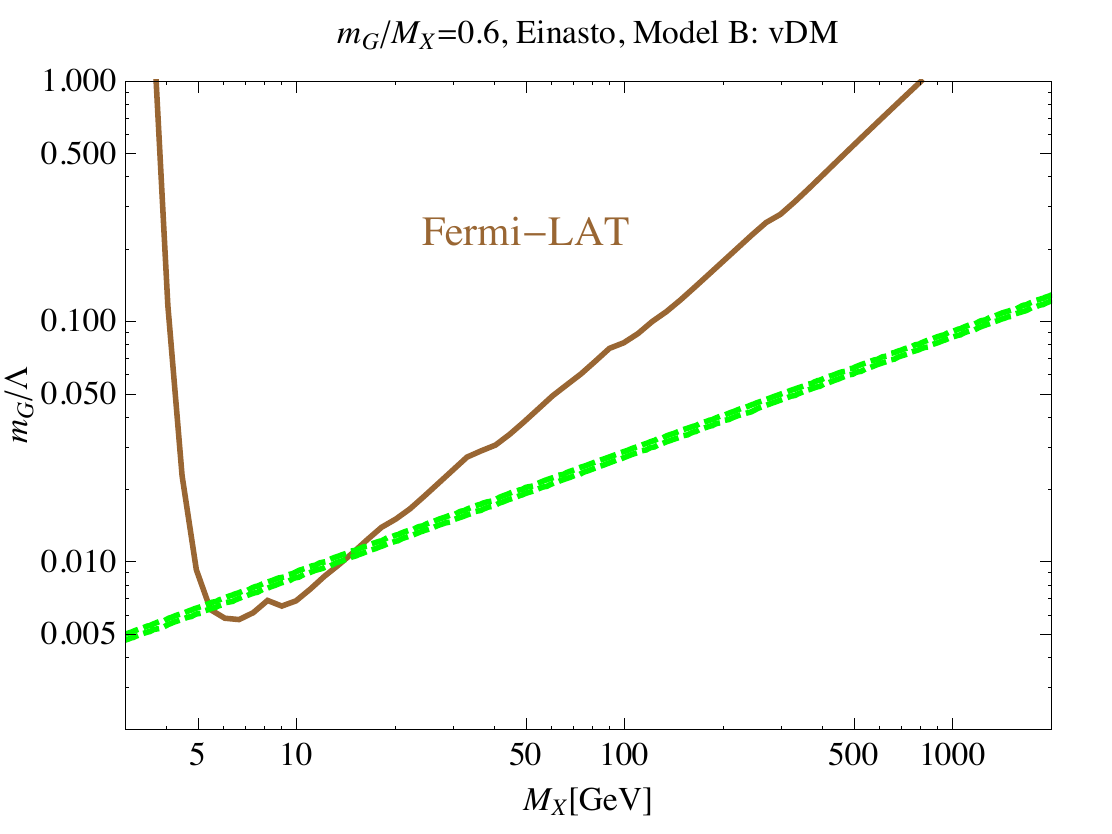}
\caption{Bounds from wide gamma-ray boxes for scalar, fermion and vector dark matter, from left to right .
We have imposed the bounds from Fermi-LAT galactic center (R16) on the annihilation cross section, $\langle\sigma v\rangle_{XX\rightarrow GG}\times {\rm Br}(G\rightarrow \gamma\gamma)$, for Einasto dark matter profile \cite{axionbox}. Green dashed lines show the Planck $5\sigma$ band for the relic density.
We have taken $m_G/M_X=0.6$, $c_X=1$, $c_V=c_g=c_\gamma=0.03$ in common, and  $c_H=1$ in the upper panel (Model A) and $c_H=0$ in the lower panel (Model B).
}
\label{fig:box2}
\end{figure}

On the other hand, in Fig.~\ref{fig:box2}, we show the Fermi-LAT bounds on wide gamma-ray boxes for scalar, fermion and vector dark matter in the models.
In all the cases, it is shown that most of the parameter space being consistent with the relic density survives the Fermi-LAT gamma-ray constraints but the region with light dark matter around $5-15\,{\rm GeV}$ has been excluded for $m_G/M_X=0.6$, due to the fact that dark matter annihilation cross section into a photon pair is sizable.

\section{Direct detection and collider bounds}

In this section, we comment on the direct detection and collider bounds on the models.

As discussed in the previous section, the interactions of dark matter to the light quarks are suppressed due to the geometric separation, but gluon interactions to dark matter can be sizable and dominant in determining the relic density for $m_{\rm DM}< m_G$. In this case, the corresponding gluon interactions are relevant for the direct detection of dark matter in underground experiments \cite{hambye,GMDM}.
For instance, even if the annihilation cross section of scalar dark matter into a gluon pair is velocity-suppressed, the DM-nucleon spin-independent scattering cross section is sizable \cite{hambye,GMDM}. So, direct detection from XENON100 \cite{xenon100} or LUX \cite{lux} can constrain the parameter space with sizable KK graviton couplings at large DM masses\footnote{For instance, for $m_G=100\,{\rm GeV}$, the strongest bound for scalar dark matter with $m_S\lesssim 300\,{\rm GeV}$ is $m_G/\Lambda\lesssim 0.2$ \cite{GMDM}.  But, spin-independent cross sections are suppressed for light dark matter masses, allowing for a sizable KK graviton coupling. More complete analysis will be done in a future work.}, complementing the gamma-ray constraints.

On the other hand, vector dark matter only can provide monochromatic photons that are compatible with Fermi gamma-ray line, when the KK graviton mediator is heavier than dark matter. In this case, we can obtain a sizable production cross section for the KK graviton in association with a monophoton at the LHC \cite{lps,zupan,cline} and its decay into a pair of dark matter may lead to a large missing energy. Then, we can impose the similar bounds on vector dark matter from monophoton searches \cite{ATLAS-mono,CMS-mono} as for axion or $Z'$-mediated dark matter \cite{lpp}.  Furthermore, if the KK graviton decay into a pair of dark matter is forbidden, we should rely on the resonant production of the KK graviton via gluon fusion into diphotons~\cite{spin2Higgs},  or vector boson or photon fusion~\cite{VBFDM}. 

Other decays of the graviton, such as top or W,Z boson pairs, would be difficult to be performed at low mass, as the current resonant searches are based on boosted topologies. Decays to Higgs pairs have been studied in Ref.~\cite{Rojo}, in a method which interpolates between the non-boosted and boosted regimes.

If the dominant decay of gravitons is to dark-matter pairs, searches using mono-photon and mono-jets in association with missing energy can be re-interpreted along the lines of Ref.~\cite{monoa}, with gluon fusion as the more likely production mechanism.

\section{Conclusions}

We have investigated a new dark matter model where gravity or composite sector such as a KK graviton in 5D warped spacetime or a spin-2 resonance in the dual conformal theory is responsible for the annihilation of dark matter into the SM particles. 
Dark matter annihilates mainly into gauge bosons, because the SM fermions and/or Higgs fields are separated from dark matter geometrically in the extra dimension.  As a result, the KK graviton coupling to dark matter can allow for the correct relic density in the perturbative regime but it tends to be larger for scalar and fermion dark matters due to the velocity-suppressed annihilation cross section than for vector dark matter.  We have shown that in the case of vector dark matter, the annihilation of dark matter into a photon pair can explain the Fermi gamma-ray line for $m_X\simeq 133\,{\rm GeV}\lesssim m_G$. 

Vector dark matter can be most strongly constrained by gamma-ray data, because the annihilation cross sections are s-wave and dark matter annihilates into photons with a sizable fraction.
In this case, most of the parameter space below the $WW$ threshold being compatible with the relic density is in tension with the current gamma-ray constraints of Fermi-LAT line search in the galactic center. On the other hand, for $m_X\gtrsim m_G$, for which the correct relic density is obtained mainly by the annihilation of dark matter into a pair of the KK gravitons, most of the parameter space is consistent with the relic density and the current gamma-ray constraints, apart from the region with the light dark matter around $10\,{\rm GeV}$.  
The scenarios of the spin-2 mediator can be tested further in various ways, by future indirect detection with gamma-ray rays, direct detection and the resonance production of the spin-2 mediator at the LHC.

\section*{Acknowledgments}
HML would like to thank CERN Theory Group for its hospitality during his visits to CERN in Jan 2013 and Jan 2014 where the present work was initiated and finally finished.
The work of HML is supported in part by Basic Science Research Program through the National Research Foundation of Korea(NRF) funded by the Ministry of Education, Science and Technology(2013R1A1A2007919). The work of MP is supported by a CERN-Korean fellowship and the World Premier International Research Center Initiative (WPI Initiative), MEXT, Japan. The work of VS is supported by the Science Technology and Facilities Council (STFC) under grant number ST/J000477/1.




\end{document}